\documentclass[12pt]{article}
\usepackage{draft_UMONS}

\usepackage{genyoungtabtikz}
\Yboxdim{8pt}
\Ylinethick{1pt}
\newcommand{\R}{\mathbb{R}}
\newcommand{\Enveloping}{\mathcal{U}}
\newcommand{\Ideal}{\mathcal{I}}

\newcommand{\Casimir}{\mathcal{C}}
\newcommand{\hs}{\mathfrak{hs}}
\newcommand{\hsdeformed}{{\ensuremath{\pmb{\mathrm{A}}_u}}}
\newcommand{\AlgInd}[1]{\mathsf{#1}}
\newcommand{\1}{\mathbf{1}}
\newcommand{\Centraliser}{\mathcal{Z}}
\newcommand{\Weyl}{\mathcal{A}}
\newcommand{\WeylClifford}{\mathcal{A}}
\newcommand{\bosOsc}{\mathsf{a}}
\newcommand{\ferOsc}{\mathsf{c}}

\newcommand{\ket}[1]{\lvert#1\rangle}
\newcommand{\Maxwell}{\mathcal{M}}
\newcommand{\Joseph}{\mathcal{J}}

\begin{document}

\title{%
    Partially-massless higher spin algebras\\
    in four dimensions
}

\author{Thomas \textsc{Basile}}
\author{Shailesh \textsc{Dhasmana}}

\affiliation{%
    Service de Physique de l'Univers, Champs et Gravitation, \\
    Universit\'e de Mons, 20 place du Parc, 7000 Mons, Belgium
}

\abstract{%
    We propose a realisation of partially-massless
    higher spin algebras in four dimensions in terms of
    bosonic {\it and} fermionic oscillators, using
    Howe duality between $sp(4,\mathbb R) \cong so(2,3)$
    and $osp(1|2(\ell-1), \mathbb R)$. More precisely,
    we show that the centraliser of $osp(1|2(\ell-1),\R)$
    in the Weyl--Clifford algebra generated by $4$ bosonic
    and $8(\ell-1)$ fermionic symbols, modulo $osp(1|2(\ell-1),\R)$
    generators, is isomorphic to the higher spin algebra
    of the type-A$_\ell$ theory whose spectrum contains
    partially-massless fields of all spins
    and depths $t=1,3,\dots,2\ell-1$. We also discuss
    the possible existence of a deformation of this algebra,
    which would encode interaction for the type-A$_\ell$ theory.
}

\maketitle

\section*{Introduction}
From a holographic perspective, theories with massless
higher spin fields in anti-de Sitter (AdS) spacetime should be
dual to free conformal field theories (CFT)
\cite{Maldacena:2011jn, Boulanger:2013zza, Alba:2013yda, Alba:2015upa}.
In all dimensions, one can distinguish between either
the free scalar or the free fermion theory, and in even dimensions,
an additional possibility exists in the guise of the free
$\tfrac{d-2}2$-form \cite{Siegel:1988gd}.
Giving up unitarity on the CFT side allows one to consider 
higher derivative versions of these theories, 
\begin{equation*}
    S[\phi] = \tfrac12\,\int_{\R^d} d^dx\,\phi^*\Box^\ell\phi\,,
    \qquad 
    S[\psi] = i\,\int_{\R^d} d^dx\,\bar\psi\,
    \slashed{\partial}^{2\ell-1}\psi\,,
\end{equation*}
where $\ell \geq 1$ is an integer, $\phi^*$ denotes
the complex conjugate of the scalar field $\phi$ and
$\bar\psi$ the Dirac adjoint of the spinor field $\psi$.
Although non-unitary, these theories capture
some non-trivial physics, namely they describe a special
type of fixed point of the renormalisation group flow, 
called the multi-critical isotropic Lifshitz points
\cite{Bekaert:2013zya, Diehl:2002ri}.
Such theories contain currents of arbitrary integer spin,
which are both conserved and partially-conserved.
More precisely, and focusing on the scalar case,
the $2\ell$-derivative scalar --- also known
as order-$\ell$ singleton --- has currents of the form
\begin{equation*}
    J_{a_1 \dots a_s}^{(t)}
    = \phi^* \partial_{a_1} \dots \partial_{a_s}
    \Box^{\ell-\frac{t+1}2}\phi + \dots\,,
\end{equation*}
for all integers $s \geq 1$ and $t=1,3,\dots,2\ell-1$,
where the dots stand for additional terms which ensure
that the above tensor is traceless and obey
the partial-conservation law
\begin{equation*}
    \partial^{b_1} \dots \partial^{b_t}
    J^{(t)}_{a_1 \dots a_{s-t} b_1 \dots b_t} \approx 0\,,
\end{equation*}
where $\approx$ signifies that this holds
when the scalar field is on-shell, i.e. $\Box^\ell\phi\approx0$
\cite{Bekaert:2013zya} (see also \cite{Brust:2016zns}).
Such currents are dual to spin-$s$ partially-massless
fields of depth-$t$ \cite{Dolan:2001ih}, which are fields
propagating an intermediary number of degrees of freedom
between those of a massless field and those of a massive
one with the same spin \cite{Deser:1983mm,Higuchi:1986wu,Deser:2001us}
(see also \cite{Deser:1983tm, Higuchi:1986py,Higuchi:1989gz,Deser:2001wx, Zinoviev:2001dt,  Deser:2001pe}). These fields can be realised as totally
symmetric rank-$s$ tensors in AdS, schematically $\varphi_{s,t}$,
subject to $t$-derivative gauge symmetry
\begin{equation*}
    \delta_\xi \varphi_{s,t} \sim \nabla^t \xi_{s-t}\,,
\end{equation*}
where $\xi_{s-t}$ is any rank-$(s-t)$ symmetric tensor
(see also \cite{Skvortsov:2006at} for their frame-like
formulation, and \cite{Krasnov:2021nsq,Basile:2022mif}
for a recently proposed twistor-inspired formulation
in four dimensions). The number of derivatives $t$
in the gauge transformations is called the `depth'
of the partially-massless, with $t=1$ corresponding
to the massless case.

Based on the spectrum of partially-conserved currents
of the order-$\ell$ singleton,
the dual higher-spin gravity,
usually referred to as the type-A$_\ell$ theory,
should contain partially-massless fields
of any integer spins and {\it odd} depths
from $1$ to $2\ell-1$, i.e. schematically
\begin{equation*}
    \text{Type-A}_\ell = \bigoplus_{t=1,3,\dots,2\ell-1}
    \bigoplus_{s=t}^\infty \varphi_{s,t} + (\cdots)\,,
\end{equation*}
where the dots stand for finitely many {\it massive} fields
(i.e. with no gauge symmetry), and of spin
lower than $2\ell-1$
(see e.g. \cite{Bekaert:2013zya, Basile:2014wua}).
The higher spin algebra is the algebra of global symmetry
of such a theories, so that its spectrum is given directly
by that of massless and partially-massless fields,
namely it is given by gauge parameters which leads to 
trivial gauge symmetries. These are known to be generalised
Killing tensors \cite{Nikitin1991},
which form finite-dimensional representations
of $so(2,d)$ corresponding to Young diagrams
with two rows, of length $s-1$ and $s-t$ respectively,
so that the type-A$_\ell$ higher spin algebra
--- denoted by $\hs_\ell$ hereafter ---
admits the decomposition
\begin{equation*}
    \newcommand{\smo}{\scalebox{0.75}{$s-1$}}
    \newcommand{\smt}{\scalebox{0.75}{$s-t$}}
    \Yboxdim{9.5pt}
    \hs_\ell \cong \bigoplus_{t=1,3,\dots,2\ell-1}
    \bigoplus_{s=t}^\infty\
    \gyoung(_5<\smo>,_3<\smt>)\,,
\end{equation*}
as an $so(2,d)$-module.
Partially-massless higher spin algebras have been previously
considered in \cite{Eastwood2008, Gover2009, Michel2014, Alkalaev:2014nsa, Joung:2015jza, Brust:2016zns},
and have been characterised both as a quotient
of the universal enveloping algebra of $so(2,d)$
by a specific ideal, and in terms of the Howe duality
between $so(2,d)$ and $sp(2,\R)$ in the Weyl algebra
generated by $2 \times (d+2)$ variables \cite{Joung:2015jza}.
Implementing these quotients may be cumbersome in practice,
even in the case $\ell=1$, i.e. for the `massless'
higher spin algebra.

The latter does admit a simple realisation
in \emph{four dimensions}, namely as the even subalgebra
of the Weyl algebra (in two pairs of oscillators).
This is due to the existence of the low-dimensional
isomorphism involving the AdS$_4$ isometry algebra,
$so(2,3) \cong sp(4,\R)$, and the fact that the module
of the ($\ell=1$) singleton is realised simply
in the Fock space associated to this Weyl algebra.
In this paper, we argue that one can simply append
a Clifford algebra (whose number of generators 
is related to $\ell$) to the Weyl algebra, and recover 
the type-A$_\ell$ algebra as a `simpler' quotient,
using Howe duality between $sp(4,\R)$
and $osp\big(1|2(\ell-1),\R\big)$. Such a realisation,
technically easier to work with, could be useful
for the introduction of interactions and the derivation
of a partially-massless higher spin gravity.

This paper is organised as follows: in Section \ref{sec:HSA},
we start by briefly reviewing the usual constructions
of the type-A$_\ell$ higher spin algebra mentioned previously,
and then proceed to present our realisation based on extending
the Weyl algebra with a Clifford algebra.
In Section \ref{sec:singleton}, we propose a construction
of the higher order singleton modules using the Fock space
naturally associated with the Weyl--Clifford algebra
used to realise the type-A$_\ell$ higher spin algebra.
We discuss in Section \ref{sec:formalHiSGRA}
the possible existence of formal partially-massless
higher spin gravities, and exhibit a cycle of the type-A$_\ell$
algebra in arbitrary dimension as part of this analysis.
We also explicitly verify that some of the Joseph ideal 
generators vanish in our oscillator realisation.
We conclude the paper in Section \ref{sec:discu}
with a discussion about the difficulties one may encounter
in construction deformations of the type-A$_\ell$ higher spin
algebra using our realisation. Appendix \ref{app:A2}
contains additional details on the structure
of the defining ideal for the type-A$_2$ algebra
(as well as some comments for the arbitrary $\ell$ case)
in arbitrary dimensions.

\section{Partially-massless higher spin algebra}
\label{sec:HSA}
\subsection{Lightning review in arbitrary dimensions}

\paragraph{Quotient of the universal enveloping algebra.}
First, let us set some notation: we will denote by $M_\AlgInd{AB}$
the generator of the Lie algebra $so(2,d)$, with indices
$\AlgInd{A,B},\dots$ taking $d+2$ values, and by $\eta_\AlgInd{AB}$
the (components of the diagonal) metric of signature $(-,-,+,\dots,+)$.
The Lie bracket of these $so(2,d)$ generators reads
\begin{equation}
    [M_\AlgInd{AB},M_\AlgInd{CD}]
    = \eta_\AlgInd{BC}\,M_\AlgInd{AD}
    -\eta_\AlgInd{AC}\,M_\AlgInd{BD}
    -\eta_\AlgInd{BD}\,M_\AlgInd{AC}
    +\eta_\AlgInd{AD}\,M_\AlgInd{BC}\,,
\end{equation}
and we will simply denote the associative product
in the universal enveloping algebra $\Enveloping\big(so(2,d)\big)$
by juxtaposition, for instance we will write
\begin{equation}
    \Casimir_2 = -\tfrac12\,M_\AlgInd{AB}\,M^\AlgInd{AB}\,,
\end{equation}
for the quadratic Casimir operator of $so(2,d)$,
where the indices have been raised with the inverse metric
$\eta^\AlgInd{AB}$. The higher spin algebra of type-A$_\ell$
is the quotient \cite{Gover2009, Michel2014}
\begin{equation}
    \hs_\ell = {\Enveloping\big(so(2,d)\big)}
    \big/{\Ideal_\ell}\,,
\end{equation}
of the universal enveloping algebra of $so(2,d)$
by the (two-sided) ideal\footnote{The notation 
$\big\langle (\cdots) \big\rangle$ means that
the ideal is generated by the elements inside
the brackets.}
\begin{equation}
    \Ideal_\ell = \Big\langle V_\AlgInd{ABCD}
    \oplus \big(\Casimir_2 + \tfrac{(d-2\ell)(d+2\ell)}4\,\1\big)
    \oplus \Joseph_{\AlgInd{A}(2\ell)}\Big\rangle\,,
\end{equation}
where 
\begin{equation}
    V_\AlgInd{ABCD} := M_\AlgInd{[AB}\,M_\AlgInd{CD]}\,,
    \qquad 
    \Joseph_{\AlgInd{A}(2\ell)} := M_\AlgInd{A}{}^\AlgInd{B_1}\,M_\AlgInd{AB_1} \dots M_\AlgInd{A}{}^\AlgInd{B_\ell}\,M_\AlgInd{AB_\ell} - \text{traces}\,,
\end{equation}
and where we used the convention
(standard in the higher spin literature)
that symmetrised indices are denoted by
the same letter, with their number being indicated
in parenthesis when necessary,
e.g. $A(l)=(A_1...A_l)$.

Recall that the universal enveloping algebra
of a Lie algebra $\mathfrak{g}$ is isomorphic,
as vector space%
\footnote{Actually as a $\mathfrak{g}$-module,
and as a (co-commutative) coalgebra.},
to the symmetric algebra $S(\mathfrak{g})$. This space
is, by definition, the symmetrised tensor product of
the adjoint representation $\gyoung(;,;)$ of $so(2,d)$,
and can be decomposed into a direct sum of finite-dimensional
irreducible representations that we will denote by
the corresponding Young diagram. In such terms,
the subspace of elements quadratic in the Lie algebra
generators reads
\begin{equation}
    \gyoung(;,;)^{\odot2} \cong \gyoung(;;,;;)
    \oplus \gyoung(;,;,;,;) \oplus \gyoung(;;)
    \oplus \bullet\,,
\end{equation}
where in particular
\begin{equation}
    \gyoung(;,;,;,;)
    \quad\longleftrightarrow\quad 
    V_\AlgInd{ABCD}
    \qquad\text{and}\qquad
    \bullet \quad\longleftrightarrow\quad \Casimir_2\,.
\end{equation}
When modding out the ideal $\Ideal_\ell$, the totally antisymmetric
diagram is removed, whereas the quadratic Casimir operator
is related to a multiple of the identity. More precisely,
the quadratic Casimir operator is set to take the value $-\tfrac14(d-2\ell)(d+2\ell)$,
which is the same value it takes when acting on the order-$\ell$
singleton module. Next we can look at the subspace of the universal
enveloping algebra spanned by elements cubic in the Lie algebra,
\begin{equation}
    \gyoung(;,;)^{\odot3} \cong \gyoung(;;;,;;;)
    \oplus \gyoung(;;,;;,;,;) \oplus \gyoung(;,;,;,;)
    \oplus \gyoung(;;,;,;) \oplus \gyoung(;;;,;)
    \oplus \gyoung(;,;)\,,
\end{equation}
and make the following observations.
\begin{enumerate}[label=$(\roman*)$]
\item First, the three diagrams
with more than two rows are contained in the product
of the ideal generators $V_\AlgInd{ABCD}$ and the Lie algebra
generators $M_\AlgInd{AB}$, and hence belong to the ideal
$\Ideal_\ell$,
\begin{equation}
    \gyoung(;,;,;,;) \oplus \gyoung(;;,;;,;,;)
    \oplus \gyoung(;;,;,;) \subset \Ideal_\ell\,,
\end{equation}
so that they are removed once $\Ideal_\ell$ is modded out 
from the universal enveloping algebra of $so(2,d)$. This is,
in fact, a general pattern: Young diagrams with more than
two rows appearing in the decomposition of
$\Enveloping\big(so(2,d)\big)$ all belong to the ideal
$\Ideal_\ell$, and more specifically, to the ideal generated
by $V_\AlgInd{ABCD}$. As a consequence, the higher spin algebra 
$\hs_\ell$ contains only Young diagrams with one or two rows.
\item Second, the diagram $\gyoung(;,;)$ is obtained
as the product of the quadratic Casimir operator $\Casimir_2$
with the Lie algebra generators $M_\AlgInd{AB}$.
Since, after modding out the ideal $\Ideal_\ell$,
the value of $\Casimir_2$ is fixed, the adjoint representation
only appears with multiplicity one in $\hs_\ell$.
\end{enumerate}
One can immediately extract from the previous item
the following lesson: in order
for the quotient algebra $\hs_\ell$ to admit
a \emph{multiplicity-free} decomposition under $so(2,d)$,
i.e. that all irreducible representations (irreps)
appear only once in the decomposition of $\hs_\ell$
under the adjoint action of $so(2,d)$,
the center of the universal enveloping algebra
has to be fixed. In other words, modding out the ideal
$\Ideal_\ell$ should fix the values of all Casimir operators
(quadratic and higher), as the latter form a basis
of the center of $\Enveloping\big(so(2,d)\big)$.

Finally, since $so(2,d)$-module appearing
in the decomposition of the universal enveloping algebra
$\Enveloping\big(so(2,d)\big)$ are contained,
by definition, in tensor product of its adjoint
representation, all these irreps are characterised by
Young diagrams with an \emph{even} number of boxes.
In view of the previous discussion, this means that
the are necessarily of the form $\gyoung(_5,_3)$,
where difference between the number of boxes
in the first and in the second row is \emph{even}.
This difference equals $t-1$ where, as before,
$t$ is the depth of the partially massless field.
Modding out by the symmetric diagram 
$\gyoung(_5{\scriptstyle 2\ell})$
effectively removes all diagrams with $t>2\ell-1$,
as they would belong to the product of the former
with another diagram in the spectrum.

\paragraph{Howe duality in the Weyl algebra.}
Now consider the Weyl algebra $\Weyl_{2(d+2)}$ generated by
$\{Y^\AlgInd{A}_i\}$ where $i=\pm$, and with the Moyal--Weyl
star-product
\begin{equation}
    f \star g = f\,\exp
    \big(\tfrac{\overleftarrow{\partial}}{\partial Y^\AlgInd{A}_i}\,
    \eta^\AlgInd{AB} \epsilon_{ij}\,
    \tfrac{\overrightarrow{\partial}}{\partial Y^\AlgInd{B}_j}\big)\,g\,,
\end{equation}
where $\epsilon_{ij}$ are the components of the canonical
$2 \times 2$ symplectic matrix, as associative product.
Quadratic monomials in $Y^\AlgInd{A}_i$, i.e. linear
combinations of the generators
\begin{equation}
    K^\AlgInd{AB}_{ij}
    := \tfrac12\,Y^\AlgInd{A}_i Y^\AlgInd{B}_j\,,
\end{equation}
span a Lie subalgebra isomorphic to $sp\big(2(d+2),\R\big)$,
\begin{equation}
    [K^\AlgInd{AB}_{ij}, K^\AlgInd{CD}_{kl}]_\star
    = \eta^\AlgInd{BC}\epsilon_{jk}\,K^\AlgInd{AD}_{il}
    + \eta^\AlgInd{AC}\epsilon_{ik}\,K^\AlgInd{BD}_{jl}
    + \eta^\AlgInd{BD}\epsilon_{jl}\,K^\AlgInd{AC}_{ik}
    + \eta^\AlgInd{AD}\epsilon_{il}\,K^\AlgInd{BC}_{jk}\,,
\end{equation}
where $[-,-]_\star$ denotes the commutator with respect to 
the star-product. The index structure on display here allows
one to easily identify two mutually commuting Lie subalgebras,
\begin{equation}
    so(2,d) \oplus sp(2,\R) \subset sp\big(2(d+2),\R\big)\,,
\end{equation}
respectively generated by
\begin{equation}
    M_\AlgInd{AB} := \tfrac12\,\epsilon^{ij}\,
    Y^\AlgInd{A}_i Y^\AlgInd{B}_j\,,
    \qquad \text{and} \qquad 
    \tau_{ij} := \tfrac12\,\eta_\AlgInd{AB}\,
    Y^\AlgInd{A}_i Y^\AlgInd{B}_j\,.
\end{equation}
Such pairs of algebras are usually called reductive dual pairs,
or Howe dual pairs \cite{Howe1989i, Howe1989ii, Goodman2000},
and can be used to construct a realisation of the type-A$_\ell$
higher spin algebra in the Weyl algebra.

To do so, we will first need to identify the centraliser 
$\Centraliser_{\Weyl_{2(d+2)}}\big(sp(2,\R)\big)$
of $sp(2,\R)$ in the Weyl algebra $\Weyl_{2(d+2)}$,
which is the space of elements annihilated by
\begin{equation}
    [\tau_{ij},-]_\star = Y^\AlgInd{A}_{(i}\,\epsilon_{j)k}\,
    \tfrac{\partial}{\partial Y^\AlgInd{A}_k}\,,
\end{equation}
or equivalently by the three operators
\begin{equation}
    Y^\AlgInd{A}_+\,
    \tfrac{\partial}{\partial Y^\AlgInd{A}_+}
    - Y^\AlgInd{A}_-\,
    \tfrac{\partial}{\partial Y^\AlgInd{A}_-}\,,
    \qquad\qquad 
    Y^\AlgInd{A}_+\,
    \tfrac{\partial}{\partial Y^\AlgInd{A}_-}\,,
    \qquad\qquad 
    Y^\AlgInd{A}_-\,
    \tfrac{\partial}{\partial Y^\AlgInd{A}_+}\,.
\end{equation}
The first operator imposes that elements in the centraliser 
of $sp(2,\R)$ be of the same degree in $Y^\AlgInd{A}_+$
and $Y^\AlgInd{A}_-$, while the other two operators
both impose that the coefficients of monomials
in $Y_\pm^\AlgInd{A}$ have the symmetry of a rectangular
Young diagram in the $so(2,d)$ indices. In other words,
\begin{equation}
    f(Y) \in \Centraliser_{\Weyl_{2(d+2)}}\big(sp(2,\R)\big)
    \qquad \Longleftrightarrow \qquad 
    f(Y) = \sum_{s=1}^\infty f_{\AlgInd{A}(s-1),\AlgInd{B}(s-1)}\,
    Y^{\AlgInd{A}(s-1)}_+\,Y^{\AlgInd{B}(s-1)}_-\,,
\end{equation}
with $f_{\AlgInd{A}(s-1),\AlgInd{AB}(s-2)} = 0$.
Note however that these tensors are still traceful,
and hence are \emph{reducible} representations of $so(2,d)$.
Decomposing them into irreducible representations,
one would find all possible finite-dimensional irreps
of $so(2,d)$ labelled by Young diagrams of the form
\begin{equation}
    \Yboxdim{11pt}
    \gyoung(_7<s-1>,_4<s-t>)
    \qquad\text{with}\qquad
    s \geq 1\,, \quad t \in 2\,\mathbb N+1\,,
\end{equation}
i.e. all Young diagrams with two rows whose lengths differ
by an even number of boxes. Comparing to the universal
enveloping algebra construction reviewed previously,
we found ourselves with the same content as we do
after modding out $\Enveloping\big(so(2,d)\big)$ by
the ideal generated by $V_\AlgInd{ABCD}$. We also
face the same multiplicity problem: recall that we need
to fix the center of the universal enveloping algebra
in order to obtain a multiplicity-free spectrum. Here, 
the source of multiplicities is not only the center
of the universal enveloping algebra of $so(2,d)$,
but also that of $sp(2,\R)$ which is, by definition,
also contained in the centraliser of $sp(2,\R)$
in $\Weyl_{2(d+2)}$. Fortunately, both problems
can be solved at once thanks to the fact that the quadratic
Casimir operators of $so(2,d)$ and $sp(2,\R)$ are related
via
\begin{equation}
    \Casimir_2[so(2,d)] + \Casimir_2[sp(2,\R)]
    = -\tfrac14\,(d-2)(d+2)\,,
\end{equation}
and similarly for higher order Casimir operator
(see e.g. \cite{Klink1988, Leung1993, Leung1994, Itoh2003} and \cite[Sec. 9]{Basile:2020gqi} for more details).
Since $sp(2,\R)$ only has one independent Casimir operator, 
it is sufficient to fix its value to also fix the values
of all Casimir operators of $so(2,d)$. In particular,
imposing
\begin{equation}
    \Casimir_2[sp(2,\R)] = -(\ell-1)(\ell+1)\,,
\end{equation}
sets the quadratic Casimir operator of $so(2,d)$ to
\begin{equation}
    \Casimir_2[so(2,d)] = -\tfrac14\,(d-2\ell)(d+2\ell)\,,
\end{equation}
as it should in the type-A$_\ell$ algebra $\hs_\ell$.
Finally, notice that the diagrams of shape $(s-1,s-t)$
with $t=2k+1$ and $k \geq 0$ appear as the $k$th trace
of rectangular diagrams, and that these traces are proportional
to $k$ times the $sp(2,\R)$ generators. As a consequence,
one can recover the partially-massless higher spin algebra
as the quotient\footnote{Note that
$\tau_{(i_1j_1} \dots \tau_{i_\ell j_\ell)}$ generate
the annihilator of the finite-dimensional
$sp(2,\R)$-irrep of highest weight $\ell-1$,
which is a reflection of the fact that the order-$\ell$
singleton is Howe dual to this $\ell$-dimensional
irrep of $sp(2,\R)$ \cite{Alkalaev:2014nsa, Joung:2015jza}.}
\begin{equation}
    \hs_\ell \cong \Centraliser_{\Weyl_{2(d+2)}}\big(sp(2,\R)\big)\Big/
    \big\langle \tau_{(i_1 j_1} \dots \tau_{i_\ell j_\ell)} \oplus
    \Casimir_2[sp(2,\R)] - \tfrac12\,(\ell-1)(\ell+1)\1\big\rangle\,,
\end{equation}
as modding out the ideal generated by the elements 
$\tau_{(i_1j_1} \dots \tau_{i_\ell j_\ell)}$ guarantees
that only $so(2,d)$ Young diagrams corresponding to
partially-massless fields of depth $t=1,3,\dots,2\ell-1$.
See e.g. \cite{Vasiliev:2004cm, Bekaert:2004qos, Bekaert:2008sa, Didenko:2014dwa, Joung:2015jza, Sharapov:2018kjz}
for more details on the construction of higher spin algebras
from the perspective of Howe duality.

\subsection{Four-dimensional specificities}
\paragraph{Dual pairs and the Weyl--Clifford algebra.}
Consider a set of bosonic $(Y^A)$ and fermionic $(\phi^A_i)$
where the capital indices $A,B,\dots$ take $2n$ values
and the lower case indices $i,j, \dots$ take $2p$ values.
These oscillators are subject to the commutation
and anticommutation relations
\begin{subequations}
    \begin{align}
        \big[\hat Y^A, \hat Y^B\big]
        & = 2\,C^{AB}\,\1\,, \label{eq:Y} \\
        \big\{\hat \phi^A_i, \hat \phi^B_j\big\}
        & = 2\,C^{AB}\epsilon_{ij}\1\,,
    \end{align}
\end{subequations}
where $C^{AB}=-C^{BA}$ and $\epsilon_{ij}=-\epsilon_{ji}$ are 
two antisymmetric, non-degenerate matrices, with inverses given by
\begin{equation}
    C^{AC}\,C_{BC} = \delta^A_B\,,
    \qquad
    \epsilon^{ik}\,\epsilon_{jk} = \delta^i_j\,.
\end{equation}
We can therefore use these matrices to raise and lower indices,
which we will do using the convention 
\begin{equation}
    C^{AB}\,X_B = X^A\,,
    \qquad
    X^A\,C_{AB} = X_B\,,
\end{equation}
and similar convention for $\epsilon_{ij}$. The associative
algebra generated by these oscillators modulo the above
anti/commutation relations forms the Weyl--Clifford algebra
$\WeylClifford_{2n|4np}$, which is simply the tensor product
of the Weyl algebra generated by the bosonic oscillators,
and the Clifford algebra generated by the fermionic ones.

The elements quadratic in these oscillators
(modulo the previous anti/commutation relations),
\begin{equation}
    K^{AB} := \tfrac14\big\{\hat Y^A, \hat Y^B\big\}\,,
    \qquad 
    M^{AB}_{ij} := \tfrac14\,\big[\hat \phi^A_i, \hat \phi^B_j\big]\,,
    \qquad 
    Q^{A|B}_i := \tfrac12\,\hat Y^A\,\hat \phi^B_i\,,
\end{equation}
form a subalgebra isomorphic to $osp(4np|2n,\mathbb R)$,
whose bosonic subalgebra $o(4np) \oplus sp(2n,\mathbb R)$
is generated by $M^{AB}_{ij}$ and $K^{AB}$,
and the odd/fermionic generators --- the supercharges
--- correspond to $Q^{A|B}_i$. Their anti/commutation
relations read
\begin{subequations}
    \begin{align}
        \big[K^{AB}, K^{CD}\big]
        & = C^{BC}\,K^{AD} + C^{AC}\,K^{BD}
        + C^{BD}\,K^{AC} + C^{AD}\,K^{BC}\,,\\
        \big[M^{AB}_{ij}, M^{CD}_{kl}\big]
        & = C^{BC}\epsilon_{jk}\,M^{AD}_{il}
        - C^{AC}\epsilon_{ik}\,M^{BD}_{jl}
        - C^{BD}\epsilon_{jl}\,M^{AC}_{ik}
        + C^{AD}\epsilon_{il}\,M^{BC}_{jk}\,,\\
        \big[K^{AB}, Q^{C|D}_i\big]
        & = C^{BC}\,Q^{A|D}_i + C^{AC}\,Q^{B|D}_i\,,\\
        \big[M^{AB}_{ij}, Q^{C|D}_k\big]
        & = C^{BD}\epsilon_{jk}\,Q^{C|A}_i
        -C^{AD}\epsilon_{ik}\,Q^{C|B}_j\,,\\
        \big\{Q^{A|C}_i, Q^{B|D_j}\big\}
        & = C^{AB}\,M^{CD}_{ij}
        + C^{CD}\epsilon_{ij}\,K^{AB}\,.
    \end{align}
\end{subequations}
Note that the orthogonal algebra is presented
in a slightly unconventional basis here: one should
think of the pair of indices $(A,i)$ on the generators
$M^{AB}_{ij}$ and $Q^{B|A}_i$ as a single index
for the fundamental representation of $o(4np)$.
This is accordance with the fact that only the first 
capital index of the fermionic generators $Q^{A|B}_i$
(the index `$A$' here) is rotated by the $sp(2n,\mathbb R)$
generators $K^{AB}$, whereas the second capital index
is rotated, along with the lower case index (the indices
`$B$' and `$i$' here) are rotated together by the $o(4np)$
generators $M^{AB}_{ij}$.

This unusual structure of indices for the $o(4np)$
generators, which stems from the choice of indices
carried out by the fermionic oscillators $\phi^A_i$,
is motivated by the fact that we are interested in
singling out the pair of subalgebras
\begin{equation}
    sp(2n, \mathbb R) \oplus sp(2p, \mathbb R) \subset o(4np)\,,
\end{equation}
generated by
\begin{equation}
    J^{AB} := \epsilon^{ij}\,M^{AB}_{ij}\,,
    \qquad \text{and} \qquad 
    \tau_{ij} := C_{AB}\,M^{AB}_{ij}\,,
\end{equation}
i.e. the generators obtained by contracting
those of $o(4np)$ with the invariant tensors
$\epsilon^{ij}$ of $sp(2p,\mathbb R)$, and
$C_{AB}$ of $sp(2n,\mathbb R)$, respectively.
As is clear from the index structure of these generators,
these two subalgebras commute with one another, 
i.e. they are contained in each other's centraliser
in $o(4np)$, and in fact they are exactly
their respective centralisers.

\paragraph{An interlude on Howe duality.}
Such pair of subalgebras are usually called `dual pairs'
and have particularly interesting applications in representation
theory and physics. The most famous examples come from dual
pairs in a symplectic group $Sp(2N,\mathbb R)$,
which are the central object of study of Howe duality 
\cite{Howe1989i, Howe1989ii} (see e.g. . In this case, one can show
that the oscillator representation of $Sp(2N,\mathbb R)$,
i.e. the Fock space generated by $N$ pairs of bosonic
creation-annihilation operators, admits a decomposition
into direct sum of the tensor product of a representation
of each group of the dual pair.

Another variation on the same theme consists in
considering dual pairs in an \emph{orthogonal} group,
say $O(2N)$.
This is precisely the case we are presented
with above, with the pair $\big(sp(2n,\R), sp(2p, \R)\big)
\subset o(4np)$. For such dual pairs, the natural
representation of the orthogonal group is the Fock space
generated by \emph{fermionic} pairs of creation-annihilation
operators. Indeed, bilinear in these operators define
a representation of the orthogonal group (or the double
cover thereof) on the fermionic Fock space, which can then
be decomposed into irreducible representations of
the dual pair of interest.
See e.g. \cite{Rowe:2011zz, Rowe:2012ym} for more details 
on this `skew-Howe' duality.

Since we have both bosonic and fermionic oscillators
at hand, we can consider dual pairs in the orthosymplectic
group \cite{Cheng2000, Cheng2010, Cheng2012}. In our case,
the relevant pair is composed of $sp(2n,\R)$, generated by
\begin{align}
   T^{AB} := K^{AB} - \epsilon^{ij}\,M^{AB}_{ij}
   = \tfrac14\,\big\{\hat Y^A, \hat Y^B\big\}
   - \tfrac14\,\big[\hat \phi^A_i, \hat \phi^{Bi}\big]\,,
\end{align}
and satisfying the commutation relations
\begin{subequations}
    \begin{align}
        [T^{AB}, \hat Y^C] &= C^{AC}\,\hat Y^B + C^{BC}\,\hat Y^A\,,\\
        [T^{AB}, \hat \phi^C_i] &= C^{AC}\,\hat \phi^B_i
        + C^{BC}\,\hat \phi^A_i\,,\\
        [T^{AB}, T^{CD}] &= C^{AC}\,T^{BD} + C^{AD}\,T^{BC}
        +C^{BD}\,T^{AC} + C^{BC}\,T^{AD}\,,
    \end{align}
\end{subequations}
and $osp(1|2p,\R)$, generated by
\begin{equation}
    Q_i = \tfrac12\,C_{AB}\,\hat Y^A\,\hat \phi^B_i\,,
    \qquad \text{and} \qquad 
    \tau_{ij} \equiv \big\{Q_i, Q_j\big\}
    = \tfrac14\,C_{AB}\,[\hat \phi^A_i, \hat \phi^B_j]\,,
\end{equation}
obeying,
\begin{subequations}
    \begin{align}
        [\tau_{ij}, Q_k] &= \epsilon_{kj}\,Q_i+\epsilon_{ki}\,Q_j\,,\\
        [\tau_{ij}, \tau_{kl}] &= \epsilon_{ki}\tau_{jl}
        + \epsilon_{kj}\tau_{il} + \epsilon_{li}\tau_{jk}
        + \epsilon_{lj}\tau_{ik}\,.
    \end{align}
\end{subequations}

\paragraph{Casimir operators.}
The quadratic Casimir operators for $sp(2n,\mathbb R)$
and $osp(1|2p,\mathbb R)$ are respectively given by,
\begin{align}
    \Casimir_2\big[sp(2n,\mathbb R)\big]
    = -\tfrac14\,T_{AB}\,T^{AB}\,,
    \qquad 
    \Casimir_2\big[osp(1|2p,\mathbb R)\big]
    = -\tfrac12\,Q_i Q^i - \tfrac14\,\tau_{ij} \tau^{ij}\,,
\end{align}
and a direct computation shows that, in the previously
described oscillator realisation, these Casimir operators
are related to one another via
\begin{equation}
    \Casimir_2\big[sp(2n,\mathbb R)\big]
    = \tfrac{n}8\,(2p-1)(2p+2n+1)
    -\Casimir_2\big[osp(1|2p,\mathbb R)\big]\,.
\end{equation}
In particular, for $n=2$ and $p=\ell-1$,
one finds
\begin{equation}
    \Casimir_2\big[sp(4,\mathbb R)\big]
    + \Casimir_2\big[osp(1|2(\ell-1),\mathbb R)\big]
    = -\tfrac14\,(3-2\ell)(3+2\ell)\,,
\end{equation}
this last number being the values of the quadratic
Casimir operator of $so(2,3) \cong sp(4, \mathbb R)$
on the module of the order-$\ell$ scalar singleton.
This is a first hint that one may recover the type-A$_\ell$
higher spin algebra as the centraliser of $osp(1|2(\ell-1),\R)$
in the Weyl--Clifford algebra, modulo $osp(1|2(\ell-1),\R)$
generators, as we shall prove in the next paragraphs.

\paragraph{Partially-massless higher spin algebra.}
In order to identify the type-A$_\ell$ higher spin algebra, 
let us first give an equivalent presentation of the Weyl--Clifford
algebra in terms of symbols of the previous oscillators,
that we will denote by $Y^A$ and $\phi^A_i$ and which are
commuting and anticommuting respectively. Their product
is the graded version of the previously discussed
Moyal--Weyl product\footnote{Note that, for a homogeneous
element $f \in \WeylClifford_{2n|4np}$ of degree $|f|$,
the left and right derivatives with respect to $Y^A$
and $\phi^A_i$ are related by
$f\,\tfrac{\overleftarrow{\partial}}{\partial Y^A}
= \tfrac{\partial}{\partial Y^A} f$ and 
$f\,\tfrac{\overleftarrow{\partial}}{\partial \phi^A_i}
= -(-1)^{|f|}\,\tfrac{\partial}{\partial \phi^A_i} f$.}
\begin{equation}
    f \star g = f\,\exp
    \big(\tfrac{\overleftarrow{\partial}}{\partial Y^A}\,
    C^{AB}\,\tfrac{\overrightarrow{\partial}}{\partial Y^B}
    +\tfrac{\overleftarrow{\partial}}{\partial \phi^A_i}\,
    C^{AB}\,\epsilon_{ij}\,
    \tfrac{\overrightarrow{\partial}}{\partial \phi^B_j}\big)\,g\,,
\end{equation}
where $f$ and $g$ are arbitrary polynomials in $Y^A$ and $\phi^A_i$.
The symbols of the $sp(2n,\R)$ and $osp(1|2p,\R)$ generators
are simply
\begin{equation}
    T^{AB} = \tfrac12\,Y^A Y^B
    - \tfrac12\,\epsilon^{ij}\,\phi^A_i \phi^B_j\,,
    \qquad 
    Q_i = \tfrac12\,C_{AB}\,Y^A \phi^B_i\,,
    \qquad 
    \tau_{ij} = \frac12\,C_{AB}\,\phi^A_i \phi^B_j\,,
\end{equation}
respectively.

Now let us characterise the centraliser of $osp(1|2p,\R)$
in $\WeylClifford_{2n|4np}$, that is the space of elements
annihilated by
\begin{equation}
    [Q_i,-]_\star = \phi^A_i\,\tfrac{\partial}{\partial Y^A}
    - \epsilon_{ij}\,Y^A\,\tfrac{\partial}{\partial \phi^A_j}\,,
\end{equation}
where $[-,-]_\star$ should be understood as the graded commutator
(i.e. $[f,g]_\star=f \star g - (-1)^{|f||g|}\,g \star f$
for homogeneous elements of the Weyl--Clifford algebra $f$ and $g$).
This condition is solved by considering any function
of the symbol of the $sp(2n,\R)$ generators $T^{AB}$,
\begin{equation}
    f(Y^A,\phi^B_i) \in \Centraliser_{\WeylClifford_{2n|4np}}\big(osp(1|2p,\R)\big)
    \qquad \Leftrightarrow \qquad 
    f(Y^A,\phi^B_i) = f(T^{AB})\,,
\end{equation}
since the symbols $T^{AB}$ are characteristics of
the first order partial differential equations
$[Q_i,f]_\star=0$. Due to the fact that the $sp(2n,\R)$
contain a piece quadratic in the anticommuting variables
$\phi^A_i$, the only possible diagram that can appear
when decomposing the centraliser of $osp(1|2p,\R)$
are those whose second row (and by extension, all rows 
except the first one) are of length smaller than $2p$.
Indeed, upon splitting the $sp(2p,\R)$ indices as
$i=(+\alpha,-\alpha)$ with $\alpha=1,\dots,p$,
we have
\begin{equation}
    \tfrac12\,\epsilon^{ij}\,\phi^A_i\,\phi^B_j
    = \sum_{\alpha=1}^p \varphi^{AB}_\alpha\,,
    \qquad\quad
    \varphi^{AB}_\alpha
    := \phi^{(A}_{+\alpha}\,\phi^{B)}_{-\alpha}\,,
\end{equation}
where
\begin{equation}
    \varphi_\alpha^{(AB}\,\varphi^{CD)}_\alpha = 0\,,
\end{equation}
by virtue of the fact that $\phi^A_i$ are anticommuting.
Note also that since the `building blocks' of the centraliser
of $osp(1|2p,\R)$ are rank-$2$ symmetric tensors of $sp(2n,\R)$,
all diagrams appearing will have an even number of boxes,
and in particular, each row will be of even length. This means
that, for $n=2$, diagrams appearing in the centraliser
of $osp(1|2p,\R)$ will be of the form
\begin{equation}\label{eq:spectrum}
    \Yboxdim{11pt}
    \gyoung(_7<2s-t-1>,_4<t-1>)\,,
\end{equation}
with $s\geq1$ and $t=1,3,\dots,2p+1$, so that upon setting
$p=\ell-1$ as before, we recover exactly the spectrum
of diagrams expected to appear in the higher spin algebra
$\hs_\ell$ (in the $sp(4,\R)$ basis). However, these diagrams
are not traceless in $sp(4,\R)$ sense a priori: consider
for instance the product of two $sp(4,\R)$ generators,
which can be projected onto a totally symmetric part,
\begin{equation}
    T^{ABCD} := T^{(AB}\,T^{CD)}
    \qquad \longleftrightarrow \qquad 
    \gyoung(;;;;)\,,
\end{equation}
and a piece with the symmetry of a `window-shaped' diagram,
\begin{equation}
    T^{AB,CD} := T^{AB}\,T^{CD} - T^{A(C}\,T^{D)B}
    \qquad \longleftrightarrow \qquad 
    \gyoung(;;,;;)\,.
\end{equation}
While the first one, the totally symmetric part,
is trivially traceless, the second one is not since
\begin{equation}
    C_{BC}\,T^{AB,CD}\
    \propto\ 2\,\epsilon^{ij}\,\phi^{[A}_i Y^{D]}\,Q_j
    - \epsilon^{ij}\epsilon^{kl}\,\phi^A_i \phi^D_k\,\tau_{jl}\,,
\end{equation}
does not vanish identically, but is proportional to 
the $osp(1|2(\ell-1),\R)$ generators. This is in fact a general
feature, namely all traces are proportional to these generators.
Indeed, taking a trace in the $sp(4,\R)$ sense
means contracting the capital latin indices $A,B,\dots$
with the invariant tensor $C_{AB}$, which thereby produces
the generators $Q_i$ and $\tau_{ij}$. Consequently, we can remove
traces by modding out the centraliser of $osp(1|2(\ell-1),\R)$
by the ideal generated by $Q_i$ and $\tau_{ij}$, and thereby
obtain the type-A$_\ell$ higher spin algebra in four dimensions
as the quotient\footnote{Note that this definition
also works for $\ell=1$, even though this case may seem degenerate
at first glance. Indeed, in this case the Howe dual algebra
becomes trivial, which is simply a consequence of the fact
the relevant Howe dual \emph{group} is the finite group
$\mathbb{Z}_2$. This group acts on the Weyl algebra 
by reflections $Y^A \to - Y^A$, so that its centraliser
is nothing but the \emph{even} subalgebra, the subalgebra
of polynomials in an even number of $Y^A$'s.}
\begin{equation}
    \hs_\ell \cong \Centraliser_{\WeylClifford_{4|8(\ell-1)}}
    \big(osp(1|2(\ell-1),\R)\big)\Big/
    \big\langle Q_i \oplus \tau_{ij}\big\rangle\,.
\end{equation}
The main difference compared to the realisation reviewed
in the previous section is that here, the `order' $\ell$
of the theory is no longer controlled by choosing different
ideal to mod out from the centraliser of the Howe dual algebra,
but by the choice of the Howe dual algebra itself. This allows
us to slightly simplify the identification of the type-A$_\ell$
higher spin algebra in four dimensions with respect to
the arbitrary dimension construction. 

Given that $\hs_\ell$ is the symmetry algebra
of the order-$\ell$ scalar singleton, and having found
a realisation of it within the Weyl–Clifford algebra,
it is natural to seek a realisation of the order-$\ell$ singleton
in the Fock space generated by the bosonic and fermionic oscillators 
used above, which we will do in the next section.

\section{Higher order singleton module}
\label{sec:singleton}
The Weyl--Clifford algebra generated by the oscillators
$\hat Y^A$ and $\hat \phi^A_i$ introduced
in Section \ref{sec:HSA} naturally acts on the Fock space
generated by $n$ pairs of bosonic creation-annihilation
operators,
\begin{equation}
    [\bosOsc_a, \bar\bosOsc^b] = \delta^b_a\,\1\,,
    \qquad 
    \bar\bosOsc^a := (\bosOsc_a)^\dagger\,,
    \qquad
    a,b,\dots=1,\dots,n\,,
\end{equation}
and $2np$ fermionic ones,
\begin{equation}
    \{\ferOsc^i_a, \bar\ferOsc_j^b\}
    = \delta^i_j\,\delta^b_a\,\1\,,
    \qquad
    \bar \ferOsc^a_i := (\ferOsc_a^i)^\dagger\,,
    \qquad 
    i,j,\dots=1,\dots,2p\,.
\end{equation}
In fact, the Weyl--Clifford algebra is the algebra
of endomorphisms of this Fock space.
Bilinears in these creation-annihilation operators
form a Lie subalgebra isomorphic to $osp(2n|4np,\R)$,
which contain the dual pair $sp
(2n,\R) \oplus osp(1|2p,\R)$
discussed previously. Introducing the notation,
\begin{equation}
    v \cdot w := \epsilon^{ij}\,v_i\,w_j
    = \epsilon_{ij}\,v^i\,w^j\,,
\end{equation}
for the contraction of the $sp(2p,\R)$ indices,
the generators of $sp(2n,\R)$ are given by
\begin{subequations}
    \begin{equation}
        T^{ab} := \bar\bosOsc^a\,\bar\bosOsc^b
        - \bar\ferOsc^a \cdot \bar\ferOsc^b\,,
        \qquad
        \qquad 
        T_{ab} := \bosOsc_a\,\bosOsc_b
        - \ferOsc_a \cdot \ferOsc_b\,,
    \end{equation}
    \begin{equation}
        T^a{}_b := \bar\bosOsc^a\,\bosOsc_b
        + \bar\ferOsc^a\cdot\ferOsc_b
        + \tfrac{1-2p}2\,\delta^a_b\,\1\,,
    \end{equation}
\end{subequations}
while the generators of $osp(1|2p,\R)$ read
\begin{equation}
    Q_i := \tfrac12\,(\bar\ferOsc^a_i\,\bosOsc_a
    - \bar\bosOsc^a\,\epsilon_{ij}\,\ferOsc_a^j)\,,
    \qquad 
    \tau_{ij} := \epsilon_{k(i}\,\bar\ferOsc_{j)}^a\,\ferOsc^k_a\,.
\end{equation}

Now let us isolate the $sp(2n,\R)$ representation
dual to the trivial irrep of $osp(1|2p,\R)$. Doing so
amounts to finding states in the Fock space
which are annihilated by the action of the $osp(1|2p,\R)$
supercharges, i.e. 
\begin{equation}
    Q_i\,f(\bar\bosOsc,\bar\ferOsc) \lvert 0 \rangle = 0\,,
\end{equation}
which is solved by 
\begin{equation}
    f(\bar\bosOsc,\bar\ferOsc) = f(T^{ab})\,,
\end{equation}
that is, any function of the $sp(2n,\R)$ raising operators
$T^{ab}$. Since the vacuum $\lvert 0 \rangle$ of the Fock space
is $osp(1|2p,\R)$-invariant, it defines a lowest weight vector
for the dual $sp(2n,\R)$-module, with weight 
\begin{equation}
    (\underbrace{\tfrac{1-2p}2,\dots,
    \tfrac{1-2p}2}_{n\,\text{times}})\,,
\end{equation}
with respect to the Cartan subalgebra spanned by
the generators $T^a{}_a$ (no summation implied).
The subspace of homogeneous polynomials of degree $k$
in $T^{ab}$ is preserved by the action of the $u(n)$ subalgebra
generated by $T^a{}_b$ (since the latter preserve
the number of creation/annihilation operators). The decomposition
of these subspaces into irreducible representations of $u(n)$
consists of all Young diagram with $2k$ boxes, whose rows
are all of even length and such that the second row is of length
at most $2p$. In particular, for $n=2$, the lowest weight 
$sp(4,\R)$-module dual to the trivial $osp(1|2p,\R)$-representation
admits the decomposition
\begin{equation}
    \mathcal{D}_{sp(4,\R)}\big(\tfrac{1-2p}2,\tfrac{1-2p}2\big)
    \cong \bigoplus_{s=0}^\infty \bigoplus_{k=0}^p
    \big[\tfrac{1-2p}2+2s+2k,\tfrac{1-2p}2+2k\big]_{u(2)}\,,
\end{equation}
under the maximal compact subalgebra $u(2) \subset sp(4,\R)$.
Taking into account the isomorphism
\begin{equation}
    [\lambda_1,\lambda_2]_{u(2)}
    \cong \big[\tfrac{\lambda_1+\lambda_2}2,
    \tfrac{\lambda_1-\lambda_2}2\big]_{so(2) \oplus so(3)}\,,
\end{equation}
between finite-dimensional irreps of $u(2)$
and $so(2) \oplus so(3)$ and setting $p=\ell-1$,
this decomposition matches the one of the order-$\ell$
singleton module
\begin{equation}
    \mathcal{D}_{so(2,3)}\big(\tfrac{3-2\ell}2,0\big)
    \cong \bigoplus_{s=0}^\infty
    \bigoplus_{t=1,3,\dots}^{2\ell-1}
    \big[\tfrac{3-2\ell}2+t-1+s,s\big]_{so(2) \oplus so(3)}
\end{equation}
in three dimensions (see e.g. \cite[Sec. 3.4.2]{Iazeolla:2008ix}).

\paragraph{A word about non-unitarity.}
Let us conclude this section by commenting
on the non-unitarity of these modules. Recall that 
higher-order singletons are lowest weight irreps
of $so(2,d)$, which can therefore be described 
by its lowest weight vector $\ket\phi$, obeying
\begin{equation}
    (D-\Delta_\phi)\ket{\phi} = 0\,,
    \qquad J_{ab}\ket\phi = 0\,,
    \qquad K_a\ket{\phi} = 0\,,
\end{equation}
where $D$, $J_{ab}$ and $K_a$ are the dilation,
Lorentz, and special conformal transformation generators.
All states of the modules are obtained
by repeated application of the translation generators
$P_a$ on $\ket\phi$. Using the relations
\begin{equation}
    [K_a, P_b] = \eta_{ab}\,D - M_{ab}\,,
    \qquad 
    [D,P_a] = P_a\,,
    \qquad 
    [M_{ab},P_c] = 2\,\eta_{c[b}\,P_{a]}\,,
\end{equation}
one finds
\begin{equation}
    K_a\,P^2\,\ket{\phi}
    = 2\,\big(\Delta_\phi-\tfrac{d-2}2\big)\,
    P_a\ket{\phi}\,,
\end{equation}
and with $P_a^\dagger = K_a$, this implies
\begin{equation}
    \|P^2\,\ket{\phi}\|^2 = 2d\,\Delta_\phi\,
    \big(\Delta_\phi-\tfrac{d-2}2\big)\,
    \langle\phi\!\mid\!\phi\rangle\,.
\end{equation}
The above identities tells us that $P^2\,\ket\phi$
is singular and null for $\Delta_\phi=\tfrac{d-2}2$,
while for $\Delta_\phi<\tfrac{d-2}2$ it is not singular
but acquires a negative norm.
For the order-$\ell$ singleton,
$\Delta_\phi=\tfrac{d-2\ell}2$ and hence the presence 
of $P^2\ket\phi$ is one of the first indications
that the module is non-unitary.

Now coming back to our construction,
it may be surprising that such a non-unitary module 
can be realised in a Fock space, which is usually 
itself a unitary module (for the Heisenberg algebra,
or its supersymmetric version relevant here).
A first consistency check is that this negative
norm state $P^2\ket{\phi}$ is indeed present,
since we recover the correct $u(2)$ decomposition.
More importantly, the Hermitian conjugation
\emph{does not preserve} the $osp(1|2p,\R)$
generators in this realisation, which is why
we have a non-unitary module in a Fock space.

\paragraph{Higher order spinor singleton?}
Note that one could look for other representation
of $osp(1|2p,\R)$ than the trivial one. For instance,
the `next-to-simplest' representation is of dimension
$2p+1 \equiv 2\ell-1$ and splits into a direct sum
of $sp(2p,\R)$ irreps, the trivial and the vector
(or fundamental) one. It can be realised
in the Fock space considered here as the subspace
with basis
\begin{equation}
    \bar\bosOsc^a\ket0 
    \qquad\text{and}\qquad
    \bar\ferOsc^a_i\ket0\,,
\end{equation}
which are indeed, for $osp(1|2p,\R)$, 
a scalar and a vector respectively.
This subspace is preserved by the action
of $osp(1|2p,\R)$ since
\begin{equation}
    Q_i\,\bar\bosOsc^a\ket0
    = \tfrac12\,\bar\ferOsc^a_i\ket0\,,
    \qquad 
    Q_i\,\bar\ferOsc_j^a\ket0
    = -\tfrac12\,\epsilon_{ij}\,
    \bar\bosOsc^a\ket0\,,
\end{equation}
while the $sp(2p,\R)$ generators $\tau_{ij}$
merely rotate this states, as expected.
As usual in the context of Howe duality, 
this representation appears with a \emph{multiplicity},
as indicated by the fact that the above basis vectors
also carry an $sp(2n,\R)$ index. In fact, 
as in the case of the trivial representation,
any state obtained from the above basis vectors
by the action of $osp(1|2p,\R)$-invariant operators,
which are generated by the Howe dual algebra $sp(2n,\R)$,
will not change the representation. In other words, 
this finite-dimensional representation of $osp(1|2p,\R)$
appears with \emph{infinite multiplicity} in the Fock space,
but this feature is merely the reflection of the fact
that it is Howe dual to a lowest weight module
of $sp(2n,\R)$, which is infinite-dimensional.

The lowest weight $sp(2n,\R)$-module in question
is induced by the lowest $u(n)$-irrep spanned by
the state $\bar\bosOsc^a\ket0$ and $\bar\ferOsc^a_i\ket0$,
and generated by the action of the raising operators $T^{ab}$.
The lowest weight reads
\begin{equation}
    \big(\tfrac{3-2p}2, \tfrac{1-2p}2, \dots, \tfrac{1-2p}2\big)\,,
\end{equation}
which, in the case of $sp(4,\R) \cong so(2,3)$,
corresponds to the lowest weight 
\begin{equation}
    \big[\tfrac{3-2p}2, \tfrac{1-2p}2\big]_{u(2)}
    = \big[2-\ell,\tfrac12\big]_{so(2) \oplus so(3)}\,,
\end{equation}
whose components are respectively the conformal weight
and spin of the spinor singleton of order $\ell$
(i.e. a free spinor $\psi$ subject to the higher order
Dirac equation $\slashed{\partial}^{2\ell-1}\psi\approx0$
as recalled in the Introduction). In other words,
we find that the higher order spinor singleton
\begin{equation}
    \mathcal{D}_{sp(4,\R)}(\tfrac{3-2p}2,\tfrac{1-2p}2\big)
    \cong \mathcal{D}_{so(2,3)}\big(2-\ell,\tfrac12\big)\,,
\end{equation}
is Howe dual to the finite-dimensional $osp(1|2p,\R)$
representation made out of the trivial and vector
$sp(2p,\R)$-irreps.

This therefore begs the question: can we find the type-B$_\ell$ 
higher spin algebra in our construction? To do so, 
one would need to quotient the centraliser of $osp(1|2p,\R)$
in the Weyl--Clifford algebra by a different ideal 
than the one generated by $osp(1|2p,\R)$. Indeed, 
we saw previously that the scalar singleton is Howe dual
to the \emph{trivial} representation of $osp(1|2p,\R)$,
and hence the full algebra is the annihilator
of this representation. The quotient by $osp(1|2p,\R)$
should be understood as the quotient by the annihilator
of this trivial representation---as recalled above
when we discussed the definition of the type-A$_\ell$
algebra in arbitrary dimensions. Having this framework
in mind, we should quotient the centraliser of $osp(1|2p,\R)$
by the annihilator of its $(2\ell-1)$-dimensional irrep
in order to obtain the type-B$_\ell$ higher spin algebra.
Schematically, this means modding out by higher powers
of the $osp(1|2p,\R)$ generators, which in turn amounts
to keeping some of the traces in the diagrams \eqref{eq:spectrum},
as may be expected to reproduce the spectrum
of the type-B$_\ell$ algebra. Such an analysis is however
beyond the scope of this paper, and we leave for potential
future work.

\section{Formal partially-massless higher spin gravity}
\label{sec:formalHiSGRA}
Having build an oscillator realisation of the higher spin
algebra $\hs_\ell$ in four dimensions, we will now use it
to try and construct an interacting theory
of partially-massless higher spin fields.

The most common way of constructing formal higher spin gravities
is to consider a gauge connection $\omega$ of the relevant
higher spin algebra $\hs$, together with a zero-form $C$
taking value in a module of this algebra
(see e.g. \cite{Vasiliev:1988sa, Vasiliev:1990vu, Vasiliev:2003ev, Bekaert:2004qos, Boulanger:2008up, Boulanger:2008kw, Alkalaev:2014nsa, Didenko:2014dwa, Grigoriev:2018wrx, Sharapov:2019vyd}).
This data is associated with the coordinates
on a $Q$-manifold, which we denote by the same symbols,
and whose (co)homological vector field $Q$ encodes
the interactions. More precisely, one is then charged
with constructing equations of motion  
\begin{align}\label{eq:formal_EOM}
    d\omega & = \mathcal{V}(\omega,\omega)
    + \mathcal{V}(\omega,\omega,C) + \dots\,, \\
    dC & = \mathcal{U}(\omega,C) + \mathcal{U}(\omega,C,C)
    + \dots\,,
\end{align}
where $\mathcal V$ and $\mathcal U$ are the component of $Q$,
and the initial data for the deformation problem reads 
\begin{align}
    \mathcal{V}(a,b) & = a \star b\,,
    & \mathcal{U}(a,u) & = a \star u - u \star \pi(a)\,,
\end{align}
where $\pi$ is an anti-involution of the higher spin algebra.
At this point it is convenient to define
${}^{\mathbb Z_2}\hs=\hs \rtimes \mathbb{Z}_2$,
where $\mathbb{Z}_2=\{1, \pi\}$. In practice,
one adds an element $k$ such that $k^2=1$
and $k \star a \star k = \pi(a)$. 

Under some fairly general assumptions, one can show that
the problem of constructing the $A_\infty$-algebra
underlying the $Q$-manifold reduces to a much simpler problem
of deforming ${}^{\mathbb Z_2}\hs$ as an associative algebra \cite{Sharapov:2018hnl, Sharapov:2018ioy, Sharapov:2018kjz, Sharapov:2019vyd}.
Moreover, often times it is easy to see that
${}^{\mathbb Z_2}\hs$ can be deformed and even construct
such a deformation, which we call $\hsdeformed$, explicitly.
Once $\hsdeformed$ is available, there is an explicit
procedure to construct all vertices. For example, 
\begin{align}\label{eq:V3}
    \mathcal{V}(a,b,u) & = \phi_1(a,b) \star \pi(u)\,,
\end{align}
where $\phi_1$ is a (Hochschild) $2$-cocycle that determines
the first order deformation of ${}^{\mathbb Z_2}\hs$
to $\hsdeformed$:
\begin{align}
    a \circ b & = a \star b + u\,\phi_1(a,b) k
    + \mathcal{O}(u^2)\,.
\end{align}
The deformed algebra $\hsdeformed$ is defined from $\hs$,
the latter being usually obtained via either one
of the following constructions:
\begin{enumerate}[label=$(\alph*)$]
\item A quotient of the universal enveloping algebra
$\Enveloping\big(so(d,2)\big)$ by a two-sided ideal $\Ideal$
(in most cases called the Joseph ideal), which corresponds
to the annihilator of a given irreducible $so(2,d)$-module,
e.g. \cite{Eastwood:2002su, Vasiliev:2003ev, Joung:2015jza, Fernando:2015tiu, Gunaydin:2016bqx, Campoleoni:2021blr};
\item Using an oscillator realisation, wherein
one embeds $so(2,d)$ and its enveloping algebra
in a Weyl(--Clifford) algebra and typically obtain $\hs$
as the quotient of the centraliser of a Howe dual
algebra, as discussed above for the type-A$_\ell$ algebra,
as well as in \cite{Vasiliev:2004cm, Bekaert:2004qos,
Govil:2013uta, Didenko:2014dwa, Govil:2014uwa, Gunaydin:2016bqx}
and references therein;
\item Via the \emph{quasi-conformal} realisation,
which consists in explicitly solving the defining relations
of the (Joseph) ideal mentioned previously, see e.g. 
\cite{Gunaydin:2007vc, Fernando:2009fq, Gunaydin:2016bqx,
Sharapov:2019pdu}. 
\end{enumerate}

It has to be noted that the above form of the vertices
is non-minimal: the equations, in general,
`couple' different spins even at the free level.
In addition, such vertices become non-local starting
from $C^2$-terms \cite{Boulanger:2015ova}. 

The first order deformation defined by the $2$-cocycle
$\phi_1$ makes its presence felt already at the free level.
Indeed, linearizing the above equations around an (A)dS$_{d+1}$
background,
\begin{equation}
    \omega_0 = e^a\,P_a + \tfrac12\,\varpi^{a,b}\,L_{ab}\,,
    \qquad 
    C_0 = 0\,,
    \qquad 
    d\omega_0 + \tfrac12\,[\omega_0,\omega_0] = 0\,,
\end{equation}
their first order in the field fluctuations
should reproduce the free field equations
for partially-massless in the frame-like formalism
\cite{Skvortsov:2006at}, whose schematic form reads
\begin{equation}
    R[\omega_1]^{a(s-1),b(s-t)}
    = e_c \wedge e_d\, C_1^{a(s-1)c,b(s-t)d}\,,
    \qquad 
    R[\omega_1]^{a(s-1-m),b(s-t-n)} = 0\,,
\end{equation}
where $R[\omega_1]^{a(s-1-m),b(s-t-n)}
= \nabla \omega_1^{a(s-1-m),b(s-t-n)} + \dots$,
with $\omega_1$ the first order fluctuations
of a $1$-form valued in $\hs_\ell$\,.
More specifically, the components of $1$-form
$\omega_1$ takes values in the finite-dimensional
representations of $so(2,d)$ labelled by the two-row
Young diagrams of the form 
\begin{equation}
    \Yboxdim{11pt}
    \gyoung(_6<s-1>,_4<s-t>)
    \qquad 
    t=1,3,\dots,2\ell-1\,, \quad s=t, t+1, \dots,
\end{equation}
which corresponds to generators of the form
\begin{equation}
    M_{\AlgInd{A}(s-1),\AlgInd{B}(s-t)}
    = \underbrace{M_\AlgInd{AB}
        \cdots M_\AlgInd{AB}}_{s-t}\,
    \underbrace{M_\AlgInd{A}{}^\AlgInd{C}\,
    M_\AlgInd{AC} \cdots M_\AlgInd{A}{}^\AlgInd{C}\,
    M_\AlgInd{AC}}_{\frac{t-1}2} + \dots\,,
\end{equation}
where the dots denote terms ensuring
that the right hand side has the symmetry
of the above Young diagram, and is traceless.
The first order fluctuation of the zero-form
takes values in a representation of $\hs_\ell$,
usually called the `twisted-adjoint representation'.%
\footnote{Although it may be more relevant
to think of it as a coadjoint module
\cite[App. B]{Skvortsov:2022syz}.}
This module of the type-A$_\ell$ algebra is defined
on the same vector space as $\hs_\ell$,
but where the latter acts via a `twisted commutator'
\begin{equation}
    \mathcal{U}(\omega_0,C_1)
    = \omega_0 \star C_1 - C_1 \star \pi(\omega_0)\,,
\end{equation}
hence the name of this representation. The zero-forms
can therefore be expanded in a basis of generators
of the (partially-massless) higher spin algebra
\cite{Iazeolla:2008ix}. Typically, the Weyl tensor
$C^{a(s),b(s-t+1)}$, for the spin-$s$
and depth-$t$ partially-massless field
is the component of $C_1$ along the generator
of $\hs_\ell$ which schematically reads,
\begin{equation}
    \underbrace{L_{ab} \dots L_{ab}}_{s-t+1}
    \underbrace{P_a \dots P_a}_{t-1} P^{2\ell-t-1}
    + \dots\,,
\end{equation}
where we separated generators of $so(2,d)$
into those of the Lorentz subalgebra $so(1,d)$,
denoted by $L_{ab}$, and the transvection
(or AdS-translation) generators denoted by $P_a$.

The low spin ($s=1$ and $2$) components
of these fluctuations are given by
\begin{equation}
    \omega_1 = A \cdot \1 + h^a\,P_a
    + \tfrac12\,\omega^{ab}\,L_{ab} + \dots\,,
    \qquad 
    C_1 = \tfrac12\,F^{a,b}\,\Maxwell_{a,b} + \dots\,,
\end{equation}
where, to keep this discussion fairly general,
we denoted by $\Maxwell_{a,b}$ the generator of $\hs$
along which one finds the Maxwell tensor,
independently of the higher spin algebra of interest.
In the type-A case, it would simply be $\Maxwell_{a,b}=L_{ab}$,
while in the type-A$_\ell$ case, it would be of the form
$L_{ab}P^{2(\ell-1)} + (\dots)$ instead. 
When comparing \eqref{eq:formal_EOM}
to the previous free equations of motion,
e.g. in the spin-$1$ sector,
\begin{equation}
    dA = e_a \wedge e_b\, F^{a,b} + \dots\,,
\end{equation}
we can deduce that $\mathcal{V}$ yields
\begin{equation}
    \mathcal{V}(P_a, P_b; \Maxwell_{c,d})
    = 2\,\eta_{a[c}\,\eta_{d]b}\,\1 + \dots\,,
\end{equation}
when evaluated on $P_a \otimes P_b \otimes \Maxwell_{c,d}$.
From \eqref{eq:V3}, we know that the dots
in the previous equation originate from the expression
\begin{equation}
    \phi_1(P_a,P_b) \star \Maxwell_{a,b}
    = 2\,\eta_{a[c}\,\eta_{d]b}\,\1 + \dots\,,
\end{equation}
In other words, the product of $\phi_1(P_a, P_b)$
and the generator that corresponds to the Maxwell tensor
must contain the unit of $\hs$. Let us note
that $\phi_1(P_a, P_b)$ is also the simplest term
to probe the deformation since $\phi_1(1,\bullet)=0$
and $\phi_1(L_{ab}, \bullet)=0$. The first condition
means that the unit is not deformed and the second one
protects Lorentz symmetry. 

Recalling that $\hs$ has an invariant trace $\tr$
(defined as the projection onto the unit),
the above condition can also be rewritten as
\begin{align}
    \tr\big(\phi_1(P_a, P_b) \star \Maxwell_{a,b}\big) \neq 0\,.
\end{align}
Since the basis of any higher spin algebra $\hs$
can be decomposed into finite-dimensional
$so(d,2)$-modules, and that the trace respects $so(d,2)$, 
different generators are orthogonal to each other.
As a result, we have to have 
\begin{align}
    \phi_1(P_a, P_b)\, & \propto\, \eta_{ab}
    + T_{ab} + \dots
    && \tr\big(T_{ab} \star \Maxwell_{a,b})
    \neq 0\,,
\end{align}
where $T_{ab}$ a generator of $\hsdeformed$
that \emph{deforms} the commutator $[P_a, P_b]$.
In other words, the generator $T_{ab}$ of this deformation 
is a multiple of the \emph{dual} of the Maxwell tensor 
generator $\Maxwell_{a,b}$. In order to define
$\hsdeformed$, one needs to define
$[P_a,P_b]=L_{ab}+ u k\,T_{ab}$
and deform the Joseph ideal accordingly.

The Maxwell tensor (and the whole decomposition)
can be found by decomposing the twisted-adjoint action
$\{P_a,\bullet\}$ of translations on $C$.
The adjoint of the Lorentz algebra
may appear with multiplicity greater than $1$
(this happens for instance in the type-B or Type-A$_\ell$,
$\ell>1$, cases). The Maxwell equations
should have the form
\begin{align}
    \nabla F^{a,b} &= h_c F^{ac,b} \\
    \nabla F^{ab,c} & \propto h^{(a} F^{b),c}
    -\tfrac1d\,h_\times\,\big(\eta^{ab}F^{c,\times}
    - \eta^{c(a} F^{b),\times}\big) + \dots\,,
\end{align}
where the first line comes from the anticommutator
$\{P_m, \Maxwell_{ac,b}\}$, where $\Maxwell_{ab,c}$
is a traceless and hook-symmetric generator of the form
$\Maxwell_{a,b}P_c + (\dots)$,  and in the second line
from $\{P_m, \Maxwell_{a,b}\}$. Most importantly, $F^{a,b}$
must not contribute anywhere else. The second equation
means that, at the algebra level, one finds
\begin{align}
    \{P_{(a}, \Maxwell_{b),c}\} & = \Maxwell_{ab,c}\,.
\end{align}
which implies that $\{P^a ,\Maxwell_{a,b}\} = 0$, 
and hence this anticommutator must be a part
of the two-sided ideal defining $\hs$.
This is indeed the case for the Type-A algebra,
whose Joseph ideal contains $\{P^m ,L_{mb}\}=0$.
For the type-A$_\ell$ case, $\Maxwell_{a,b}$ must be
in the adjoint representation that sits inside
the subspace of monomials of order $\ell-1$
in the $so(2,d)$ generator (i.e. one degree less
than the generator $\Joseph_{\AlgInd{A}(2\ell)}$
of the Joseph ideal).

\paragraph{Probing deformation through cycles.}
Cocycles are more complicated than cycles to derive
since cocycles are defined on the whole algebra
(must be assigned some value for all possible arguments),
while cycles involve few specific elements of the algebra.
Nontrivial cocycles can be evaluated on nontrivial cycles,
the result being nonzero. 
In the type-A case, the Maxwell equation
probes the cycle \cite[App. B]{Sharapov:2018kjz}
\begin{equation}
    c_{(1)} = L_{ab} \otimes P^a \otimes P^b
    +\tfrac14\,(\1 \otimes L_{ab} \otimes L^{ab})
    - \tfrac14\,C_L(\1 \otimes \1 \otimes \1)\,,
    \qquad
    C_L = -\tfrac{d(d-2)}4\,,
\end{equation}
which is closed by virtue of the fact that
$\{L_{ab}, P^b\} \sim 0$
and $-\tfrac12\,L_{ab} L^{ab} \sim C_L\,\1$
due to the quotient by the Joseph ideal.
As it turns out, one can find a counterpart of this cycle
in $\hs_\ell$, namely
\begin{equation}
    c_{(\ell)} = \Maxwell_{a,b} \otimes P^a \otimes P^b
    + \tfrac12\,\big(\1 \otimes \Maxwell_{a,b} \otimes L^{ab}
    - P^a \otimes P^b \otimes \Maxwell_{a,b}
    - P^a \otimes \Maxwell_{a,b} \otimes P^b\big)
    + \dots\,,
\end{equation}
which is closed as a consequence of the fact that
$\{\Maxwell_{a,b}, P^b\} \sim 0$, \emph{up to a term}
$\1 \otimes \Maxwell_{a,b} L^{ab}$ (hence the dots).
To verify that this is indeed a cycle,
first note that
\begin{equation}
    \partial(\Maxwell_{a,b} \otimes P^a \otimes P^b)
    = \{\Maxwell_{a,b},P^a\} \otimes P^b
    - \tfrac12 \Maxwell_{a,b} \otimes L^{ab}
    \sim - \tfrac12 \Maxwell_{a,b} \otimes L^{ab}\,,
\end{equation}
as we have previously argued that $\{\Maxwell_{a,b},P^a\}$
belongs to the defining ideal of $\hs_\ell$
(in fact, any higher spin algebra
containing a massless spin-$1$ fields
in its spectrum), and hence is modded out.
The remaining term is compensated thanks to
\begin{equation}
    \partial(\1 \otimes \Maxwell_{a,b} \otimes L^{ab})
    = \Maxwell_{a,b} \otimes L^{ab}
    - \1 \otimes \Maxwell_{a,b}\,L^{ab}
    + L^{ab} \otimes \Maxwell_{a,b}\,,
\end{equation}
which however brings in two other terms. The last one
can be cancelled using
\begin{equation}
    \partial(P^a \otimes P^b \otimes \Maxwell_{a,b}
    + P^a \otimes \Maxwell_{a,b} \otimes P^b)
    \sim L^{ab} \otimes \Maxwell_{a,b}\,,
\end{equation}
where we made use of $\{\Maxwell_{a,b},P^b\}\sim0$ again,
which leaves us with the final task of eliminating
the term proportional to $\1 \otimes \Maxwell_{a,b}\,L^{ab}$.
In the type-A example, we could take advantage
of the fact that the term $L_{ab}L^{ab}$ is proportional
to the identity in $\hs$. This is a simple consequence
of quotienting $\Enveloping\big(so(2,d)\big)$
by the Joseph ideal.\footnote{More precisely,
the scalar component of the generator $\Joseph_{AB}$,
when decomposed under the Lorentz algebra, relates $P^2$
to the quadratic Casimir operator of $so(2,d)$
which is itself proportional to the identity.
Since $\Casimir_2 = -\tfrac12\,L_{ab}L^{ab} + P^2$,
one therefore concludes that $-\tfrac12\,L_{ab}L^{ab}$
is also proportional to the identity.}
We can expect that a similar property also holds
for type-A$_\ell$ algebras, by inspecting its spectrum:
since $\Maxwell_{a,b} = L_{ab}P^{2(\ell-1)} + (\dots)$
belongs to the $so(2,d)$-irrep $(2\ell-1,1)$,
the contraction $\Maxwell_{a,b}\,L^{ab}$
belongs to $(2\ell)$,
\begin{equation}
    \Yboxdim{11pt}
    \Maxwell_{a,b} \in \gyoung(_7{2\ell-1},;)
    \qquad\Longrightarrow\qquad
    \Maxwell_{a,b}\,L^{ab} \in \gyoung(_8{2\ell})
    \subset \Ideal_\ell\,,
\end{equation}
since the latter is the only $so(2,d)$ diagram
susceptible to contain a Lorentz scalar.
In other words, $\Maxwell_{a,b}\,L^{ab}$
is related to the scalar part of the generator
$\Joseph_\AlgInd{A(2\ell)}$, whose structure
is discussed in Appendix \ref{app:A2}.
We can expect that $\Maxwell_{a,b}\,L^{ab}$
is proportional to $P^{2(\ell-1)}$, or a polynomial
in $P^2$ of degree $\ell-1$ more generally.

This is indeed the case for the type-A$_2$ algebra,
where $\Maxwell_{a,b}\,L^{ab} \sim \#\,P^2$,
as we show in Appendix \ref{app:A2}. We can therefore
use this identity and compensate the term
$\1 \otimes \Maxwell_{a,b}\,L^{ab}$
by adding
\begin{equation}
    \1 \otimes P_a \otimes P^a
    \qquad\Longrightarrow\qquad
    \partial\big(\1 \otimes P_a \otimes P^a\big)
    = -\1 \otimes P^2\,,
\end{equation}
which, when added with the proper coefficient
to $c_{(2)}$ above, defines a cycle of $\hs_2$.

\paragraph{Oscillator realisation for type-A$_2$.}
Let us compute the Maxwell generator
in our oscillator realisation. To do so,
first recall that the Lorentz generators 
are embedded in $sp(4,\R)$ as
\begin{equation}
    L^{\alpha\beta}
    = \tfrac14\,\{\hat y^\alpha, \hat y^\beta\}
    - \tfrac14\,\epsilon^{ij}\,
    [\hat \phi^\alpha_i, \hat\phi^\beta_j]\,,
    \qquad 
    L^{\alpha'\beta'}
    = \tfrac14\,\{\hat y^{\alpha'}, \hat y^{\beta'}\}
    - \tfrac14\,\epsilon^{ij}\,
    [\hat \phi^{\alpha'}_i, \hat\phi^{\beta'}_j]\,,
\end{equation}
where we split the oscillators \eqref{eq:Y}
as $\hat Y^A=(\hat y^\alpha, \hat y^{\alpha'})$
with $\alpha,\alpha' \in \{1,2\}$ indices
for two-components spinors. Similarly,
the transvection generators read
\begin{equation}
    P^{\alpha\alpha'}
    = \tfrac14\,\{\hat y^{\alpha}, \hat y^{\alpha'}\}
    - \tfrac14\,\epsilon^{ij}\,
    [\hat \phi^{\alpha}_i, \hat\phi^{\alpha'}_j]\,.
\end{equation}
Let us also introduce the notation
\begin{equation}
    q_i = \tfrac12\,\hat y_\alpha\,\hat \phi^\alpha_i\,,
    \qquad 
    \bar q_i = \tfrac12\,\hat y_{\alpha'}\,
    \hat \phi^{\alpha'}_i\,,
    \qquad
    t_{ij} = \tfrac14\,\epsilon_{\alpha\beta}\,
    \big[\hat\phi_i^\alpha, \hat\phi_j^\beta\big]\,,
    \qquad 
    \bar t_{ij} = \tfrac14\,\epsilon_{\alpha'\beta'}\,
    \big[\hat\phi_i^{\alpha'}, \hat\phi_j^{\beta'}\big]\,,
\end{equation}
in terms of which the $osp(1|2p,\R)$ generators read
\begin{equation}
    Q_i = q_i + \bar q_i\,,
    \qquad 
    \tau_{ij} = t_{ij} + \bar t_{ij}\,.
\end{equation}
Note that $q_i$ and $t_{ij}$ form an $osp(1|2p,\R)$
algebra, and $\bar q_i$ and $\bar t_{ij}$ as well.
The square of the translation generators
can be written as
\begin{equation}\label{eq:P2}
    P^2 = -\tfrac12\,P_{\alpha\alpha'}\,P^{\alpha\alpha'}
    = \tfrac{2p-1}2 + q_i\,\bar q^i
    + \tfrac12\,t_{ij}\,\bar t^{ij}\,, 
\end{equation}
where the factor $-\tfrac12$ comes from
the $\gamma$-matrices used to convert vector indices
into spinor ones.\footnote{This can also be check by
comparing $[L_{ab}, P^b]$ and $[P_a, P^2]$
with their spinor counterparts, which shows that
one should use
$L_a \to \tfrac12\,\big(\epsilon_{\alpha\beta}
L_{\alpha'\beta'}+\epsilon_{\alpha'\beta'}
L_{\alpha\beta}\big)$ and
$P_a \to \tfrac{i}{\sqrt2}\,P_{\alpha\alpha'}$.}

The Maxwell generator for $p=1 \Leftrightarrow \ell=2$ 
therefore becomes
\begin{equation}
    \Maxwell_{\alpha\beta} = L_{\alpha\beta}\,
    \big(q_i\,\bar q^i + \tfrac12\,\bar t_{ij}\,t^{ij}
    +\tfrac12\big)\,,
\end{equation}
and similarly for $\Maxwell_{\alpha'\beta'}$,
upon exchanging $L_{\alpha\beta}$ with $L_{\alpha'\beta'}$.
A direct computation leads to
\begin{equation}
    \big\{\Maxwell_{\alpha\beta},P^\beta{}_{\alpha'}\big\}
    \sim 0\,,
\end{equation}
upon using the identities
\begin{equation}\label{eq:useful}
    \hat\phi^\alpha_i\,t^2
    = -2\,\hat\phi^\alpha_j\,t_i{}^j\,,
    \qquad
    t_{ik}\,t_j{}^k = -2\,t_{ij}
    +\tfrac12\,\epsilon_{ij}\,t^2\,,
    \qquad\text{and}\qquad
    (t_{ij}+2\epsilon_{ij})\,t^2 = 6\,t_{ij}\,,
\end{equation}
with $t^2 \equiv t_{ij}\,t^{ij}$,
which can be proved thanks to Fierz identities.

Let us conclude this section by pointing a subtlety
in the computation of the ideal generators
in our oscillator realisation. Introducing $\hbar$
in the canonical anti/commutation relations as
\begin{equation}
    [\hat Y^A, \hat Y^B] = 2\hbar\,C^{AB}\,,
    \qquad 
    \{\hat\phi^A_i, \hat\phi^B_j\}
    = 2\hbar\,C^{AB}\,\epsilon_{ij}\,,
\end{equation}
the $osp(1|2p,\R)$ anti/commutation relations read
\begin{equation}
    \{Q_i, Q_j\} = \hbar\,\tau_{ij}\,,
    \qquad 
    [\tau_{ij}, Q_k] = 2\hbar\,\epsilon_{k(i}\,Q_{j)}\,,
    \qquad 
    [\tau_{ij}, \tau_{kl}]
    = \hbar\,\big(\epsilon_{kj}\,\tau_{il} + \dots\big)\,,
\end{equation}
i.e. the right hand side of any anti/commutator
is proportional to $\hbar$. Contracting the second
relation with $\epsilon^{jk}$ yields
\begin{equation}
    \hbar\,q_i = -\tfrac1{2p+1}\,[t_{ij}, Q^j]
    \qquad\Longrightarrow\qquad
    \hbar^2\,t_{ij} = -\tfrac1{2p+1}\,
    \big\{[t_{ij}, Q^j], q_j\big\}\,,
\end{equation}
which could, in the absence of $\hbar$, lead one
to conclude that $q_i$ and $t_{ij}$ can be set to zero
(and similarly for $\bar q_i$ and $\bar t_{ij}$),
when taking the quotient by $osp(1|2p,\R)$. This would
however be incorrect since it would amount to quotienting
by $osp(1|2p,\R) \oplus osp(1|2p,\R)$, one copy
generated by $q_i$ and $t_{ij}$, and another copy by
$\bar q_i$ and $\bar t_{ij}$. This direct sum
is Howe dual to $sp(2,\R) \oplus sp(2,\R)$,
and not to $sp(4,\R)$, as each copy of $osp(1|2p,\R)$
does not commute with the transvection generators
$P_{\alpha\alpha'}$. Consequently, it would be inconsistent
to mod out $q_i$ and $\bar q_i$ separately (and similarly
for $t_{ij}$ and $\bar t_{ij}$) in the centraliser
of $sp(4,\R)$ --- in the sense that the resulting algebra
would not be related to the type-A$_\ell$ higher spin
algebra. 

The introduction of $\hbar$ in computation also proves
useful when it comes to checking that the scalar generator
of the ideal also vanishes: the expression \eqref{eq:P2}
of $P^2$ can be re-written as
\begin{align}
    -\tfrac12\,P_{\alpha\alpha'}\,P^{\alpha\alpha'}
    = \hbar^2\,\tfrac{2p-1}2 + q_i\,\bar q^i
    + \tfrac12\,t_{ij}\,\bar t^{ij}
    = \hbar^2\,\tfrac{2p-1}2 + q_i\,(Q^i - q^i)
    + \tfrac12\,t_{ij}\,(\tau^{ij} - t^{ij})\,,
\end{align}
Evaluating $\Joseph_\bullet^{(2)}$, which involves
the previous equation for $p=1$ and modulo $Q_i$
and $\tau_{ij}$, yields
\begin{equation}
    \Joseph_\bullet^{(2)} = \big(P^2-\tfrac{\hbar^2}2\big)\,
    \big(P^2-\tfrac{5\hbar^2}2\big)
    \sim \big(q_i\,q^i + \tfrac12\,t_{ij}\,t^{ij}\big)\,
    \big(q_k\,q^k + \tfrac12\,t_{kl}\,t^{kl}
    +2\,\hbar^2\big)
    \sim \big(q_i\,q^i + \tfrac12\,t_{ij}\,t^{ij}\big)^2\,,
\end{equation}
upon using $\hbar\,q_i \sim 0 \sim \hbar^2\,t_{ij}$.
Using again Fierz identity and \eqref{eq:useful},
one can show that 
\begin{equation}
    \big(q_i\,q^i + \tfrac12\,t_{ij}\,t^{ij}\big)^2 \sim 0
    \qquad\text{i.e.}\qquad
    \Joseph_\bullet^{(2)} \sim 0\,,
\end{equation}
modulo $Q_i$ and $\tau_{ij}$, as required.

\section{Discussion}
\label{sec:discu}
In this paper, we proposed a new realisation
of the type-A$_\ell$ higher spin algebra
in four dimensions, based on extending 
the Weyl algebra with a Clifford algebra.
This allows for an arguably simpler realisation
of $\hs_\ell$, wherein the limit of the range
of values of the depth of the partially-massless fields
is constrained by the dimension of the Clifford algebra.
We also exhibited a Hochschild $3$-cycle
of $\hs_\ell$, which suggests that there should exist
non-trivial deformations of the partially-massless
higher spin algebras.

Unfortunately, the usual technique used 
to construct deformation of higher spin algebra
that consists in using deformed oscillators,
i.e. trading $\hat Y^A$ for $\hat q^A$
which satisfy
\begin{equation}
    [\hat q^A, \hat q^B] = 2\,C^{AB}\,(\1+k\,\nu\big)\,,
\end{equation}
where $k$ is the generator of the $\mathbb Z_2$
action on the Weyl algebra,
discussed in the previous section, does \emph{not}
seem to work: we were unable to use this deformation
and preserve a realisation of the $osp(1|2p,\R)$ algebra
\emph{undeformed} while keeping the Lorentz subalgebra
undeformed as well. This situation seems surprising
since in the case of the type-B algebra, whose realisation
is also based on a quotient of the Weyl--Clifford algebra
\cite{Vasiliev:2004cm}, and are known to admit deformations
of this type \cite{Grigoriev:2018wrx, Sharapov:2019vyd}.
The deformed oscillator algebra, which first appeared
in a paper of Wigner \cite{Wigner1950},
is one of the simplest example
of a \emph{symplectic reflection algebra}
(originally introduced by Etingof and Ginzburg
\cite{Etingof2000}, see also 
\cite{Gordon2007, Bellamy2012, Chlouveraki2013}
for more recent reviews). The algebras are deformations
of the smash product of the Weyl algebra
with a finite group (acting on it by automorphisms).
The latter naturally contains reductive dual pairs
$(\mathfrak{g},\mathfrak{g}')$ of bosonic type,
which can --- at least in some cases
\cite{Feigin:2014yha, DeBie2009, Ciubotaru2018} --- 
be deformed by finding
a realisation of one of the algebra of the pair,
say $\mathfrak{g}$, in a symplectic reflection algebra.
Typically, the other algebra $\mathfrak{g}'$
is deformed to an \emph{associative} (not Lie) algebra.
In any case, both algebras are mutual centralisers
of one another, and hence one again finds a bijection
between their representations (appearing in the appropriate
Fock space). Recently, some examples of dual pairs
of Lie \emph{superalgebras} have been deformed
\cite{Ciubotaru2020, Calvert2022} using symplectic
reflection algebras. The difference with respect to 
the pair $\big(sp(2n,\R), osp(1|2p,\R)\big)$ of interest
for us is that the superalgebra $osp(1|2p,\R)$
that we would like to preserve when using deformed
oscillator has its bosonic subalgebra $sp(2p,\R)$
realised using only fermionic oscillators
which are not deformed (since they generate
a Clifford algebra which is finite-dimensional,
it does not admit a non-trivial deformation).
This seems to be one of the reason why preserving
$osp(1|2p,\R)$ appears impossible, at least
if we simply replace the bosonic oscillators $\hat Y^A$
by deformed ones $\hat q^A$ in our realisation.
We hope to come back to this issue in the near future.

\section*{Acknowledgments}
We are indebted to Zhenya Skvortsov for suggesting
this project and its main idea in the first place,
for early collaboration and continued discussions
during the completion of this work,
as well as for valuable comments on an earlier version
of this paper. We are also grateful to Euihun Joung
for useful discussions. The work of S.D. was supported
by the European Research Council (ERC)
under the European Union’s Horizon 2020 research
and innovation programme (grant agreement
No 101002551). The work of T.B. was supported by
the European Union’s Horizon 2020 research
and innovation programme
under the Marie Sk\l{}odowska Curie grant agreement
No 101034383.

\appendix 
\section{More on the type-\texorpdfstring{A$_2$}{} algebra}
\label{app:A2}
Any higher spin algebra whose spectrum consists of
totally symmetric fields \emph{only}, and defined as
a quotient of $\Enveloping\big(so(2,d)\big)$
by an ideal $\Ideal$, will necessarily contain
the antisymmetric generator\footnote{The factor
$4$ in $V_\AlgInd{ABCD}$ has been added for simplicity.}
\begin{equation}
    V_\AlgInd{ABCD} = 4\,M_\AlgInd{[AB}\,M_\AlgInd{CD]}
    = M_\AlgInd{[AB}\,M_\AlgInd{C]D}
    -M_\AlgInd{[AB}\,\eta_\AlgInd{C]D}\,,
\end{equation}
in its defining ideal $\Ideal$.
We will therefore start this appendix
by reviewing how factoring out $V_\AlgInd{ABDC}$
relate all Casimir operators to the quadratic one
(see also \cite[Sec. 2.1]{Iazeolla:2008ix}
and \cite{Boulanger:2011se}). Let us illustrate
this mechanism in the case of the quartic
Casimir operator, defined as\footnote{%
More generally, we follow the convention
that the Casimir operator of $so(2,d)$
of order $2n$ is given by
$\Casimir_{2n} := \tfrac12\,M_\AlgInd{A_1}{}^\AlgInd{A_2}\,
M_\AlgInd{A_2}{}^\AlgInd{A_3}\,\dots\,
M_\AlgInd{A_{2n}}{}^\AlgInd{A_1}$.}
\begin{equation}
    \Casimir_4 = \tfrac12\,M_\AlgInd{A}{}^\AlgInd{B}\,
    M_\AlgInd{B}{}^\AlgInd{C}\,
    M_\AlgInd{C}{}^\AlgInd{D}\,M_\AlgInd{D}{}^\AlgInd{A}\,,
    \qquad 
    \Casimir_2= -\tfrac12\,M_\AlgInd{AB}\,M^\AlgInd{AB}\,.
\end{equation}
A direct computation yields\footnote{%
For all computations in this appendix,
one needs to use a few identities
that are specific to orthogonal algebra,
which we will list here. For $so_N$,
with generators $R_{IJ}=-R_{JI}$ obeying
$[R_{IJ}, R_{KL}] = \eta_{JK}\,R_{IL} + (\dots)$
with $\eta$ of \emph{arbitrary signature}, one has
$$[R_I{}^\bullet,R_{J\bullet}]=-(N-2)\,R_{IJ}\,,
\qquad [V^\bullet, R_{I\bullet}]=-(N-1)\,V_I\,,
\qquad R_I{}^J\,R_J{}^K\,R_K{}^I
=-\tfrac{N-2}2\,R_{IJ}\,R^{IJ}\,,$$
where $V_I$ is any vector of $so_N$.}
\begin{equation}
    V_\AlgInd{ABC}{}^\bullet\,M_\AlgInd{D\bullet}
    = M_\AlgInd{AB}\,M_\AlgInd{C}{}^\bullet\,M_\AlgInd{D\bullet}
    + 2\,M_\AlgInd{C[A}\,M_\AlgInd{D}{}^\bullet\,
    M_\AlgInd{B]\bullet} + M_\AlgInd{AB}\,M_\AlgInd{CD}
    -2\,(d-1)\,M_\AlgInd{C[A}\,M_\AlgInd{B]D}\,,
\end{equation}
which, upon taking a trace in $\AlgInd{CD}$
and contracting with $M_\AlgInd{AB}$, gives
\begin{equation}\label{eq:C4-C2}
    V_\AlgInd{ABCD} \sim 0 
    \qquad\Longrightarrow\qquad
    \Casimir_4 \sim \Casimir_2\,\big(\Casimir_2
    + \tfrac{d(d-1)}2\big)\,,
\end{equation}
in agreement with \cite[Sec. 2.1]{Iazeolla:2008ix}
in the special case of the singleton,
and with \cite{Dolan:2011dv} in general.
Similarly, taking the Lorentz components $V_{abcd}$,
and contracting them with $L^{ab}$ (on the left)
and $L^{cd}$ (on the right), one finds
\begin{equation}\label{eq:C4_Lor}
    V_{abcd} \sim 0
    \qquad\Longrightarrow\qquad
    \Casimir_4^L \sim \big(\Casimir_2-P^2\big)
    \Big(\Casimir_2-P^2 + \tfrac12\,(d-1)(d-2)\Big)\,,
\end{equation}
where $\Casimir_4^L = \tfrac12\,L_a{}^b\,L_b{}^c\,
L_c{}^d\,L_d{}^a$ is the quartic Casimir operator
of the Lorentz subalgebra. This is the same type
of relation as \eqref{eq:C4-C2} with $d \to d-1$,
upon using the fact the quadratic Casimir operator
$\Casimir_2^L$ of $so(1,d)$
is given in terms of that of $so(2,d)$ by
$\Casimir_2^L = \Casimir_2 - P^2$.
Contracting $V_\AlgInd{ABCD}$ with more generators
produces similar identities,
relating Casimir operators of order $2n$
to lower order ones, and ultimately to $\Casimir_2$.

When decomposing the generator $V_\AlgInd{ABCD}$
under $so(1,d)$, one finds an additional antisymmetric
generator of rank $3$, namely $V_{abc0'}$.
Contracting it with $L^{ab}$ (on the left)
and $P^c$ (on the right) yields the identity
\begin{equation}\label{eq:LLPP}
    V_{abc0'} \sim 0
    \qquad\Longrightarrow\qquad
    L_a{}^\bullet\,L_{b\bullet}\,\{P^a,P^b\}
    \sim -2\,\big(\Casimir_2-P^2\big)\,
    \Big(P^2 + \tfrac{d-1}2\big)\,,
\end{equation}
which will be useful for us latter on.

\paragraph{Type-A$_2$.}
Let us define 
\begin{equation}
    W_\AlgInd{AB}
    := M_\AlgInd{(A}{}^\AlgInd{C}\,M_\AlgInd{B)C}\,,
\end{equation}
and consider the symmetric generator for the ideal
defining the partially-massless higher spin
algebra $A_2$, which is the traceless part
of $W^\AlgInd{(AB}\,W^\AlgInd{CD)}$, given by,
\begin{equation}
    \begin{aligned}
        \Joseph_\AlgInd{ABCD}
        & := W_\AlgInd{(AB}\,W_\AlgInd{CD)}
        -\tfrac4{d+6}\,\eta_\AlgInd{(AB}\,
        \big(W_\AlgInd{C}{}^\AlgInd{M}\,W_\AlgInd{D)M}
        - \Casimir_2\,W_\AlgInd{CD)}\big)\\
        & \hspace{120pt} + \tfrac{4}{(d+4)(d+6)}\,
        \eta_\AlgInd{(AB}\,\eta_\AlgInd{CD)}\,
        \big(\Casimir_4 + \Casimir_2\,
        [\Casimir_2 - (\tfrac{d}2)^2]\big)\,,
    \end{aligned}
\end{equation}
where we used the relation
\begin{equation}
    \tfrac12\,W_\AlgInd{AB}\,W^\AlgInd{AB}
    = \Casimir_4 - (\tfrac{d}2)^2\,\Casimir_2\,,
\end{equation}
relating the contraction of the generator $W_{AB}$
with itself and the quadratic and quartic Casimir
operators. Note that we can also express
this generator of the ideal $\Ideal_2$ as
\begin{equation}
    \Joseph_\AlgInd{ABCD}
    = \Joseph_\AlgInd{(AB}\,\Joseph_\AlgInd{CD)}
    - \tfrac4{d+6}\,\eta_\AlgInd{(AB}\,
    \Joseph_\AlgInd{C}{}^\bullet\,\Joseph_\AlgInd{D)\bullet}
    + \tfrac4{(d+4)(d+6)}\,\eta_\AlgInd{(AB}\eta_\AlgInd{CD)}\,
    \big(\Casimir_4 - \tfrac2{d+2}\,\Casimir_2^2\,
    -(\tfrac{d}2)^2\,\Casimir_2\big)\,,
\end{equation}
where
\begin{equation}
    \Joseph_\AlgInd{AB}
    := M_\AlgInd{(A}{}^\AlgInd{C}\,M_\AlgInd{B)C}
    +\tfrac2{d+2}\,\eta_\AlgInd{AB}\,\Casimir_2\,,
\end{equation}
is the traceless part of $W_\AlgInd{AB}$,
which is also one of the generator of the Joseph ideal
of the type-A algebra, and where we used
\begin{equation}
    \tfrac12\,\Joseph_\AlgInd{AB}\,\Joseph^\AlgInd{AB}
    = \Casimir_4 - \tfrac2{d+2}\,\Casimir_2^2\,
    -(\tfrac{d}2)^2\,\Casimir_2\,.
\end{equation}

This generator can be decomposed
under the Lorentz subalgebra, and in particular
contains a scalar piece,
\begin{equation}
    \Joseph^{(2)}_\bullet = \tfrac{d+2}{d+6}\,P^4
    + \tfrac4{(d+6)}\,
    \big(\tfrac14\,\{L_{ab},P^b\}\,\{L^{ac},P_c\}
    - \Casimir_2\,P^2\big) + \tfrac4{(d+4)(d+6)}\,
    \big(\Casimir_4
    + \Casimir_2[\Casimir_2-(\tfrac{d}2)^2]\big)\,,
\end{equation}
which can be re-written in terms of $\Casimir_4$,
$\Casimir_2$, $P^4$ and $P^2$, using some previously
discussed results. To do so, notice first that
\begin{equation}
    \tfrac14\,\{L_{ab},P^b\}\{L^{ac},P_c\}
    = \tfrac12\,L_a{}^\bullet\,L_{b\bullet}\,\{P^a,P^b\}
    + \tfrac{d+1}2\,(\Casimir_2-P^2)
    - \tfrac{d^2}4\,P^2\,,
\end{equation}
The first term on the right hand side can be eliminated
using \eqref{eq:LLPP},
and using the relation \eqref{eq:C4-C2}
between $\Casimir_4$ and $\Casimir_2$,
as well as imposing $\Casimir_2 \sim -\tfrac14(d-4)(d+4)$,
we end up with
\begin{equation}
    \Joseph_\bullet^{(2)}
    = \big(P^2 + \tfrac{d-4}2\big)
    \big(P^2 + \tfrac{d-8}2\big)\,.
\end{equation}

The symmetric generator $\Joseph_{\AlgInd{A}(2\ell)}$
of the defining ideal for the type-A$_\ell$
higher algebra verifies
\begin{equation}
    [M_\AlgInd{AB}, \Joseph_{\AlgInd{C}(2\ell)}]
    = 4\,\ell\,\eta_\AlgInd{C[B}\,
    \Joseph_{\AlgInd{A]C}(2\ell-1)}\,,
\end{equation}
by definition. Decomposing this identity
under the Lorentz subalgebra yields
\begin{equation}
    [P_a, \Joseph^{(\ell)}_{b(2\ell-k)}]
    = (2\ell-k)\,\eta_{ab}\,\Joseph^{(\ell)}_{b(2\ell-k-1)}
    + k\,\Joseph^{(\ell)}_{ab(2\ell-k)}\,,
    \label{eq:rec_V}
\end{equation}
for $k=0,\dots,2\ell$. We can use the above equation
to express the various Lorentz generators, obtained
by decomposing $\Joseph_{\AlgInd{A}(2\ell)}$,
in terms of the scalar one
\begin{equation}\label{eq:J0}
    \Joseph^{(\ell)}_\bullet
    = \sum_{k=0}^\ell \nu_{2k}(\Casimir_{2n})\,P^{2k}\,,
\end{equation}
where $\nu_k$ are polynomials in the Casimir
operators of $so(2,d)$, and $\nu_{2\ell}=1$.
Indeed, for $k=2\ell$ the equality \eqref{eq:rec_V}
yields
\begin{equation}
    \Joseph^{(\ell)}_a
    = \tfrac1{2\ell}\,[P_a, \Joseph_\bullet^{(\ell)}]\,,
\end{equation}
while for $k=2\ell-1$ it gives,
\begin{equation}
    \Joseph^{(\ell)}_{ab} = \tfrac1{2\ell\,(2\ell-1)}\,
    \big[P_a, [P_b, \Joseph_\bullet^{(\ell)}]\big]
    + \tfrac1{(2\ell-1)}\,\eta_{ab}\,\Joseph_\bullet^{(\ell)}\,,
\end{equation}
which can then be used to obtain, recursively,
expressions for all generators $\Joseph^{(\ell)}_{a(2\ell-k)}$
given by various linear combinations of nested
commutators of $P_a$ and $\Joseph^{(\ell)}$. Schematically,
\begin{equation}
    V_{a(k)}^{(\ell)} = \sum_{j=0}^{[k/2]} \#\,
    \underbrace{\eta_{aa} \dots \eta_{aa}}_{j\,\text{times}}\,
    [\underbrace{P_a, \dots, [P_a}_{k-2j\,\text{times}},
    \Joseph^{(\ell)}_\bullet] \dots]\,,
\end{equation}
where $\#$ generically denotes combinatorial coefficients
that can be obtained by recursion. For instance,
for the $\ell=1$ case, i.e. the usual type-A
higher spin algebra, the symmetric generator
\begin{equation}
    \Joseph_\AlgInd{AB}
    = M_\AlgInd{(A}{}^\AlgInd{C}\,M_\AlgInd{B)C}
    + \tfrac2{d+2}\,\eta_\AlgInd{AB}\,\Casimir_2\,,
\end{equation}
decomposes into three generators,
\begin{equation}
    \Joseph^{(1)}_{ab} = L_{(a}{}^c\,L_{b)c} - P_{(a}\,P_{b)}
    + \tfrac2{d+2}\,\eta_{ab}\,\Casimir_2\,,
    \quad 
    \Joseph^{(1)}_a = \tfrac12\,\{L_{ab}, P^b\}\,,
    \quad 
    \Joseph^{(1)}_\bullet = P^2 -\tfrac2{d+2}\,\Casimir_2\,,
\end{equation}
and one can check that the rank-$2$ symmetric
and the vector generators can be re-written as
\begin{equation}
    \Joseph^{(1)}_{ab} = \tfrac12\,\big[P_a, [P_b, P^2]\big]
    + \eta_{ab}\,(P^2 - \tfrac2{d+2}\,\Casimir_2)\,,
    \qquad 
    \Joseph^{(1)}_a = \tfrac12\,[P_a, P^2]\,.
\end{equation}
For $\ell=2$, one finds
\begin{equation}
    \Joseph^{(2)}_a = \tfrac12\,\Big\{L_{ab}\,
    \big(P^2 + d-3\big), P^b\Big\}\,,
\end{equation}
which is similar to the $\ell=1$ case, 
in that it is given by the anticommutator of $P^b$
with a monomial of order $2\ell-1=3$ in generators,
which is an antisymmetric Lorentz tensor.
In light of the discussion
in Section \ref{sec:formalHiSGRA}, 
the generator in $\tfrac12\,(L_{ab}P^2 + d-3)$
can be identified as the Maxwell generator
in type-A$_2$. In fact, this pattern holds
for arbitrary values of $\ell$:
a simple recursion leads to 
\begin{equation}
    [P_a,P^{2k}] = \sum_{j=1}^k (2-\delta_{j,k})\,
    d^{j-1}\,\big\{L_{ab}\,P^{2(k-j)},P^b\big\}\,,
    \qquad k\geq1\,,
\end{equation}
which yields
\begin{equation}
    \Joseph_a^{(\ell)}
    = \tfrac1{2\ell}\,\sum_{k=0}^{\ell-1}
    a_{2k}\,\big\{L_{ab}\,P^{2k}, P^b\big\}\,,
    \qquad\text{with}\qquad
    a_{2k} = (2-\delta_{k,0})
    \sum_{j=k+1}^\ell d^{j-k-1}\,\nu_{2j}\,,
\end{equation}
where $\nu_{2j}$ denote the coefficients
in the expression of $\Joseph_\bullet^{(\ell)}$
as a polynomial in $P^2$ \eqref{eq:J0}.

\newpage
\footnotesize

\providecommand{\href}[2]{#2}\begingroup\raggedright\endgroup

\end{document}